\documentstyle[12pt,epsf,epsfig]{article}
 \oddsidemargin  = - 7 mm 
 \evensidemargin = - 7 mm
\def\tstrut{\vrule height2.5ex depth0pt width0pt} % used in tables
\newcommand{\reales}{{\rm R}\hspace{-1ex}\rule{0.1mm}{1.5ex}\hspace{1ex}}

\def\er#1#2{\relax\ifmmode{}^{+#1}_{-#2}\else$^{+#1}_{-#2}$\fi}
\newcommand{\be}{\begin{equation}}
\newcommand{\bea}{\begin{eqnarray}}
\newcommand{\ee}{\end{equation}}
\newcommand{\eea}{\end{eqnarray}}

\def\({\Big(}
\def\){\Big)}
\def\slashchar#1{\setbox0=\hbox{$#1$}
   \dimen0=\wd0 \setbox1=\hbox{/} \dimen1=\wd1
   \ifdim\dimen0>\dimen1 \rlap{\hbox to \dimen0{\hfil/\hfil}} #1
   \else  \rlap{\hbox to \dimen1{\hfil$#1$\hfil}} / \fi}
\def\P{\slashchar{P}}
\begin{document}
\title{$S=-1$ Meson-Baryon Unitarized Coupled
  Channel Chiral Perturbation Theory 
and the $S_{01}-$ $\Lambda$(1405) and
  $-\Lambda$(1670) Resonances } 
  \author{C. Garc\'{\i}a--Recio\footnote{email:g\_recio@ugr.es},
  J. Nieves\footnote{email:jmnieves@ugr.es}, E. Ruiz
  Arriola,\footnote{email:earriola@ugr.es}\\
  {\small Departamento de
  F\'{\i}sica Moderna, Universidad de Granada,
  E-18071 Granada, Spain}\\
~\\
   M. J. Vicente Vacas\\
 {\small Dpto. Fisica Te\'orica and IFIC, Centro mixto Universidad de
  Valencia-CSIC,}\\ {\small Aptd. 22085, E-46071 Valencia, Spain}\\  }

\date{\today}
\maketitle 

\thispagestyle{empty}

\begin{abstract}
  
  The $s-$wave meson-baryon scattering is analyzed for the strangeness
  $S=-1$ and isospin $I=0$ sector in a Bethe-Salpeter coupled channel
  formalism incorporating Chiral Symmetry. Four channels have been
  considered: $\pi \Sigma$, $\bar K N$, $\eta \Lambda$ and $K \Xi$.
  The required input to solve the Bethe-Salpeter equation is taken
  from lowest order Chiral Perturbation Theory in a relativistic
  formalism. There appear undetermined low energy constants, as a
  consequence of the renormalization of the amplitudes, which are
  obtained from fits to the $\pi\Sigma\to\pi\Sigma$ mass-spectrum, to
  the elastic $\bar K N \to \bar K N$ and $ \bar K N\to \pi \Sigma$
  $t$--matrices and to the $ K^- p \to \eta \Lambda$ cross section
  data. The position and residues of the complex poles in the second
  Riemann Sheet of the scattering amplitude determine masses, widths
  and branching ratios of the $S_{01}-$ $\Lambda$(1405) and
  $-\Lambda$(1670) resonances, in reasonable agreement with
  experiment. A good overall description of data, from $\pi \Sigma$
  threshold up to 1.75 GeV, is achieved despite the fact that three-body
  channels have not been explicitly included.

\end{abstract}

%\vspace*{\fill}\begin{center}Submitted to: {\it }\end{center}

\vspace*{1cm}
\centerline{\it PACS: 11.10.St;11.30.Rd; 11.80.Et; 13.75.Lb;
14.40.Cs; 14.40.Aq\\}
\vspace*{1cm} \centerline{\it Keywords: Chiral Perturbation Theory,
Unitarity, $\pi \Sigma$-Scattering, $\bar K N$-Scattering, $\eta
\Lambda$-Scattering, } 
\centerline{\it $S_{01}-\Lambda$ Resonances,
Coupled channels, Bethe-Salpeter Equation.}

%\vspace*{\fill}
\newpage
%\footnotesize
%\twocolumn
\setcounter{page}{1}

%%%%%%

\section{Introduction}

Baryon resonances are outstanding features in elastic and inelastic
meson-baryon scattering and signal the onset of non-perturbative
physics.  Constituent quark model approaches describe them as excited
baryonic bound states, and the coupling to the continuum is obtained
by evaluating transition matrix elements~\cite{AM00} but comparison
with data can only be done once the scattering problem is solved. In
such a scheme, the underlying quark constituent nature of hadrons is
taken into account but implementation of Chiral Symmetry (CS) becomes
difficult. In the region of low energies, it seems appropriate to
start considering the hadrons as the relevant degrees of freedom where
CS not only proves helpful to restrict the type of interactions
between mesons and baryons, but also provides an indirect link to the
underlying Quantum Chromo Dynamics (QCD)~\cite{Pich95}.  For processes
involving Baryons and Mesons, Heavy Baryon Chiral Perturbation Theory
(HBChPT)~\cite{JM91,BK92} incorporates CS at low energies in a
systematic way, and has provided a satisfactory description of $\pi N$
scattering in the region around threshold~\cite{Mo98,fms98,fm00}. It
suffers, however, from known limitations. Firstly, the expansion is
manifestly not relativistically invariant, and some convergence
problems, specially for the scalar form factor, have been pointed out
and solved by defining a suitable regularization scheme, the so-called
infrared regularization (IR) ~\cite{bl01}. The IR scheme has also
successfully been applied to $\pi N$ elastic
scattering~\cite{BL01}. The previous remarks concern a theory with
only $\pi $ and $N$ present.  The $\Delta$ degrees of freedom have
been explicitly included~\cite{Fettes:2000bb} within the HBChPT
scheme, where the nucleon-$\Delta$ splitting is considered to be of
order of the pion mass, pointing toward a better convergence for the
$\pi N$ scattering data than the case without explicit $\Delta$'s.
More recently, the work of Ref.~\cite{Torikoshi:2002bt} has compared
the HBChPT scheme to the IR one in the presence of explicit $\Delta$'s
degrees of freedom for $\pi N$ scattering, showing that HBChPT
describes data up to much higher pion CM kinetic energies than the IR
method.

A second limitation of any of the previous approaches, both in the
HBChPT as well as in the IR frameworks and with or without explicit
$\Delta$, is that they are based on a perturbative expansion of a
finite amount of Feynman diagrams. This complies with unitarity order
by order in the expansion, but fails to fulfill exact unitarity of the
scattering amplitude. Thus, some non-perturbative resummation should
be supplemented to incorporate exact unitarity and, hopefully, to
accommodate resonances. Regarding this second limitation, several
unitarization methods have been suggested in the literature and
previously used to describe the meson-baryon dynamics: Inverse
Amplitude Method (IAM)~\cite{pn00}, or a somehow modified IAM to
account for the large baryon masses~\cite{ej00c}, dispersion
relations~\cite{JE01,Meissner:1999vr,OM01}, Lippmann-Schwinger
Equation (LSE) and Bethe-Salpeter Equation
(BSE)~\cite{KSW95,KSW95b,Kaiser:1996js,Caro00,OR98,ORB02,Na00,Inou,JE01b,LK02}.
	
In this work, we will study the $s-$wave meson-baryon scattering for
the strangeness $S=-1$ and isospin $I=0$ sector in a Bethe-Salpeter
coupled channel formalism incorporating CS. This reaction provides a
good example of the need of unitarization methods; the recent work of
Ref.~\cite{Ka01} shows that HBChPT to one loop fails completely
already at threshold. The $\bar K N$ scattering length turns out to
have a real part about the same size but with opposite sign and half
the imaginary expected experimentally~\cite{Ma81}, due to the the
nearby subthreshold $\Lambda (1405) $-resonance. It is important to
realize from the very beginning that, for this particular resonance,
the unitarization program implies at least considering two channels,
namely $\pi \Sigma$ and $\bar K N$. Thus, one deals with a coupled
channel problem. Unfortunately, there are no one-loop ChPT
calculations incorporating this coupled channel physics. Much of the
discussion below and the rest of the paper reflects this lack of
knowledge on the general structure of this scattering amplitude within
ChPT, particularly on the number of undetermined parameters.  Of
course, not all counterterms are independent, as demanded by crossing
and chiral symmetries. Actually, the best practical way to impose
these constraints on a unitarized partial wave amplitude is by
matching to a ChPT amplitude. The reason for this is that 
the left-cut partial wave analytical structure implied by
crossing would be automatically incorporated within a one-loop ChPT
calculation.  This point of view has been used in several
unitarization approaches for the elastic $\pi N$
process~\cite{Meissner:1999vr,pn00,ej00c,JE01}. There, the ChPT
amplitude is known up to one-loop~\cite{Mo98,fms98,fm00,BL01} which in
HBChPT corresponds to third and fourth order. This is the lowest order
approximation incorporating the perturbative unitarity correction
which is required to match the perturbative amplitude to a unitarized
one. As we have already mentioned, the coupled channel ChPT one-loop
amplitude for meson-baryon scattering is not known, and thus the
matching is not possible.

The first study of the strangeness $S=-1$ and isospin $I=0$
meson-baryon channel incorporating CS and coupled channel
unitarization was carried out in Ref.~\cite{KSW95b,Kaiser:1996js},
though some phenomenological form factors were employed. More
recently, this channel has been studied in Refs.~\cite{OR98}
and~\cite{ORB02}, where a three-momentum cut-off is used to
renormalize the LSE and the off-shell behavior is partially taken into
account.  In principle, the minimal renormalization procedure used in
Refs.~\cite{OR98,ORB02} is acceptable but may turn out to be too
restrictive in practice. If, instead, one takes advantage of the
flexibility allowed by the renormalization of an Effective Field
Theory (EFT), there is a chance of improving the description from
threshold up to higher energies. In practical terms
this means increasing the number of counter terms which one has to add
to make finite the amplitudes.  Along the lines of Ref.~\cite{JE01b},
we will adopt this viewpoint below by taking fully into account the
off-shellness suggested by the BSE using the tree level amplitude as
the lowest order approximation to the potential. In the approach of
Ref.~\cite{JE01b} there are three parameters for each channel. 
In the absence of a one-loop ChPT coupled channel
amplitude, we content ourselves with matching to the tree level one
\footnote{This point was discussed at length in Ref.~\cite{EJ99} in
the context of elastic $\pi\pi$ scattering; from there one easily
realizes that the renormalization scheme pursued in Ref.~\cite{OR98}
would lead to a scenario where only one of the four
$SU(2)-$Gasser-Leutwyler $\bar l_1, \bar l_2,\bar l_3$ and $\bar l_4$
parameters is independent.}. One could also match the coupled channel
unitarized amplitude to HBChPT, much below the inelastic
thresholds. This procedure was investigated in Ref.~\cite{JE01b} for
the $\pi N$ system, but did not introduced any powerful constraints. 

Despite of the great success of the model of Ref.~\cite{OR98} in
describing the data around the antikaon-nucleon threshold, including
the features of the $S_{01}-$ $\Lambda$(1405) resonance, its
predictions for higher energies~\cite{ORB02} do not work so well and
clear discrepancies with data appear. Indeed, the limitations of the
model of Ref.~\cite{OR98} did already appear in the strangeness $S=0$
and isospin $I=1/2$ meson-baryon sector, where the model is able to
describe data only in a more or less narrow energy window around the
$N(1535)$ resonance~\cite{Na00,Inou}.  However, the model previously
developed by two of us for the latter channel in Ref.~\cite{JE01b}
describes data in a wider energy region, ranging from $\pi N$
threshold up to almost a Center of Mass (CM) meson-baryon energy of
$\sqrt{s}=2$ GeV, including the features of a second resonance
($N(1650)$). Motivated by these encouraging results we extend in this
work the model of Ref.~\cite{JE01b} to the strangeness $S=-1$ and
isospin $I=0$ meson-baryon channel.  Like in the $S=0$ sector, taking
into account the off-shellness of the BSE generates a rich structure
of unknown constants which allow for a better description of the
amplitudes.  Although the generation of more undetermined constants
may appear a less predictive approach than putting a cut-off (one
single parameter) to regularize the divergent integrals, it reflects
the real state of the art of our lack of knowledge on the underlying
QCD dynamics. The number of adjustable Low Energy Constants (LEC's)
should not be smaller than those allowed by the symmetry; this is the
only way both to falsify all possible theories embodying the same
symmetry principles and to widen the energy interval which is being
described. Limiting such a rich structure allowed by CS results in a
poor description of experimental data. However, a possible redundancy
of parameters is obviously undesirable but may be detected through
statistical considerations (see below). The number of LEC's is controlled to any order of the
calculation by crossing symmetry. In a unitarized approach, the only
way to avoid this parameter redundancy is to match the unitarized
amplitude to the one obtained from a Lagrangian formalism\footnote{See
discussion in Appendix D of Ref.~\cite{JE01b}}.  There is no standard
one loop ChPT calculation for the meson-baryon reaction with open
channels to compare with.  Some results there exist within the HBChPT
scheme, up to order ${\cal O} (1/ M^3 f^2_\pi , 1/M f^4_\pi) $ --
being $M$ a typical baryon mass -- but only involving pions and
nucleons~\cite{fm00}. An indirect way to detect such a parameter
redundancy might be through a fit to experimental data if the errors
and correlations in some parameters turn out to be very large. We will
adopt this point of view in this work, and will show that indeed
correlations take place effectively reducing the total number of
independent parameters.

In this paper four coupled channels have been considered: $\pi
\Sigma$, $\bar K N$, $\eta \Lambda$ and $K \Xi$ and taken into account
$SU(3)-$breaking symmetry effects but neglected the considerably
smaller isospin violation ones. We also neglect three body channels,
mainly for technical reasons. According to Ref.~\cite{Ma02} the
$\Lambda (1670)$ resonance decays into the $\Sigma^* \pi N$ three body
final state with a small ($ .08 \mp .06 $) branching ratio.

Preliminary results of the present work
can be found in~\cite{GR02a,GR02b,GR02c}.

The paper is organized as follows: In Sect,~\ref{sec:thf}, a summary
of the theoretical framework used is described. Further details can be
found in Ref.~\cite{JE01b}. Numerical results for amplitudes and cross
sections and details of the fitting procedure are given in
Sect.~\ref{sec:nr}. A special attention is paid both to the analytical
properties, in the complex plane, of the found $t$ matrix  and to the
statistical correlations between the fitted LEC's, being the
Appendices \ref{sec:app-stat} and \ref{app:monster} devoted to these
issues as well. Finally, in Sect.\ref{sec:concl} we outline the
conclusions of our work.  

\section{Theoretical framework} \label{sec:thf}
The coupled channel scattering amplitude for the baryon-meson process
in the isospin channel $I=0$
\begin{eqnarray} 
B( M_A , P-k,s_A ) + M( m_A , k ) \to B( M_B , P-k', s_B) + M(
m_B , k' )
\end{eqnarray}
with baryon (meson) masses $M_A$ and $M_B$ ( $m_A$ and $m_B$ ) and
 spin indices (helicity, covariant spin, etc...) $s_A, s_B$,
 is given by
\begin{equation} 
T_P \left [ B \{k',s_B\}  \leftarrow 
 A \{k,s_A\}  \right  ] = \bar u_B ( P-k', s_B) t_P (k,k') u_A
 (P-k,s_A)\label{eq:deftpeque} 
\end{equation} 
Here, $u_A (P-k, s_A)$ and $u_B (P-k', s_B)$ are baryon Dirac
 spinors\footnote{We use the normalization ${\bar u }u = 2M$.} for the
 ingoing and outgoing baryons respectively, $P$ is the conserved total
 four momentum and $t_P (k,k') $ is a matrix in the Dirac and coupled
 channel spaces. On the mass shell the parity and Lorentz invariant
 amplitude $t_P$ can be written as:
\begin{equation}
t_P (k,k') |_{\rm on-shell} = t_1 (s,t) \P + t_2 (s,t) 
\label{eq:amp-dirac}
\end{equation} 
with $s= P^2 = \P^2 $, $t = (k-k')^2$ and $t_1$ and $t_2$ matrices in
the coupled channel space.

In terms of the matrices $t_1$ and $t_2$ defined in
Eq.~(\ref{eq:amp-dirac}), the $s-$wave coupled-channel matrix,
$f_0^\frac12 (s)$ [$f_L^J$],  is  given by:
\begin{eqnarray}
\left[ f_0^\frac12 (s) \right]_{B \leftarrow A}&=& -\frac{1}{8\pi\sqrt s}
\sqrt{\frac{|\vec{k}_B|}{|\vec{k}_A|}} \sqrt{E_B + M_B}\sqrt{E_A +
M_A} \left [ \frac12 \int^1_{-1}
d\cos\theta 
\left ( \sqrt{ s} \,t_1(s,t) + t_2(s,t) \right )_{BA} \right ]
\label{eq:deff0}
\end{eqnarray}
where the CM three--momentum moduli read 
\begin{eqnarray}
|\vec{k}_i| &=& \frac{\lambda^\frac12 (s,M_i^2,m_i^2)}{ 2\sqrt {s}}\qquad i=A,B
\end{eqnarray}
with $\lambda(x,y,z) = x^2+y^2+z^2 -2xy-2xz-2yz$ and $E_{A,B}$ the
baryon CM energies. The phase of the matrix $T_P$ is such that the
relation between the diagonal elements ($A=B$) in the coupled channel
space of $f_0^\frac12 (s)$ and the inelasticities ($\eta$) and
phase-shifts ($\delta$) is the usual one,

\begin{eqnarray} 
\left [ f_0^\frac12 (s)\right]_{AA}  = {1\over 2 {\rm i} |\vec{k}_A| } 
\Big( \eta_A (s) e^{2
{\rm i} \delta_A (s)} - 1 \Big) 
\end{eqnarray}

Further details on normalizations and definitions of the amplitudes
can be seen in Section IIB of Ref.~\cite{JE01b}.

%%   Fig. 1   %%%%%%%%%%%%%%%%%%%%%%%%%%%%%%%%%%%%%%%%%%%%%%%%%%%%%
\begin{figure}[tbp]
   \centering
   \footnotesize
   \epsfxsize = 13cm
   \epsfbox{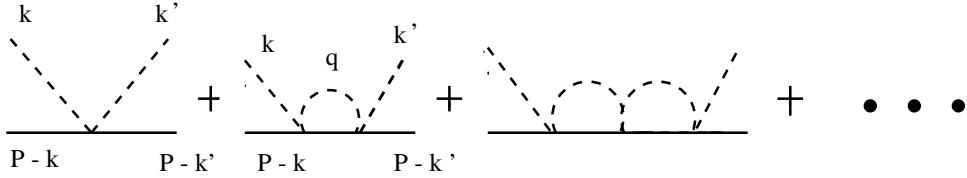} 
   \vspace*{1cm}

\caption{ \footnotesize Diagrams summed by the Bethe Salpeter
   equation. Kinematics defined in the main text.}  
\label{fig_BS}
\end{figure}
%%   Fig. 1   %%%%%%%%%%%%%%%%%%%%%%%%%%%%%%%%%%%%%%%%%%%%%%%%%%%%%

To compute the amplitude $t_P$ we solve the BSE (see Fig.~(\ref{fig_BS}))
\begin{equation} 
t_P ( k,k') = v_P ( k,k') + {\rm i} \int { d^4 q \over (2\pi)^4 }
t_P (q,k') \Delta(q) S(P-q) v_P (k,q) \label{eq:bse}
\end{equation}
where $t_P( k,k')$ is the scattering amplitude defined in
Eq.~(\ref{eq:deftpeque}), $v_P(k,k')$ the two particle irreducible
Green's function (or {\it potential} ), and $ S(P-q)$ and $\Delta (q)
$ the baryon and meson exact propagators respectively. The above
equation turns out to be a matrix one, both in the coupled channel and
Dirac spaces. For any choice of the {\it potential} $v_P(k,k')$, the
resulting scattering amplitude $t_P( k,k')$ fulfills the coupled
channel unitarity condition, discussed in Eq.~(21) of
Ref.~\cite{JE01b}.  The BSE requires some input potential and baryon
and meson propagators to be solved. To compute the lowest order of the
BSE-based expansion~\cite{EJ99} is enough to approximate the iterated
{\it potential} by the chiral expansion lowest order meson-baryon
amplitudes in the desired strangeness and isospin channel, and the
intermediate particle propagators by the free ones (which are diagonal
in the coupled channel space). From the meson-baryon chiral
Lagrangian~\cite{Pich95} (see Sect. IIA of Ref.~\cite{JE01b}), one
gets at lowest order for the {\it potential}:
\begin{equation}
v_P (k,k') = t_P^{(1)} (k,k') = {D \over f^2} ( \slashchar{k}+\slashchar{k}' ) 
\end{equation}
with $D$ the coupled-channel matrix,
\begin{eqnarray} 
\nonumber \\ 
\matrix{  \bar K N     \, \, & \quad \pi \Sigma \, \, \,& \quad  \eta
\Lambda  & \quad K \Xi  & \quad \qquad \qquad  } \nonumber  
\\ 
D^{I=0}_{S=-1} = {1\over 4} \left( \matrix{
 -3  &   \sqrt{3/2}  &   -3/\sqrt{2} & 0 \cr 
 \sqrt{3/2}  &    -4   & 0  &   -\sqrt{3/2} \cr
 -3/\sqrt{2}  &   0  &  0  &   3/\sqrt{2} \cr
 0  &   -\sqrt{3/2}  &  +3/\sqrt{2}  &  -3  
} \right)            
\qquad \matrix{   \bar K N \cr  \pi \Sigma \cr   \eta \Lambda \cr  K \Xi } 
\label{eq:d-matrix} 
\end{eqnarray} 
given in the isospin basis and the same phase conventions as in
Ref.~\cite{JE01b}.

While amplitudes follow the chiral symmetry breaking pattern from the
effective Lagrangian to a good approximation, it is well known that
physical mass splittings have an important influence when calculating
the reaction phase space. Indeed, the correct location of reaction
thresholds requires taking physical masses for the corresponding
reaction channels. We have taken into account this effect in our
numerical calculation. We also incorporate explicit CS effects to the
weak meson decay constants and different numerical values for $f_\pi$
, $f_K$ and $f_\eta$ can be used. This can be easily accomplished
through the prescription
\begin{eqnarray}
D/f^2 \to \hat f^{-1} D \hat f^{-1} \quad , \qquad \hat f \equiv {\rm
Diag} \left( f_K , f_\pi, f_\eta , f_K  \right)
\label{eq:f-presc}
\end{eqnarray}  
The solution of the BSE with the kernel specified above can
be found in Ref.~\cite{JE01b}. 
It turns out that the functions $t_1$ and $t_2$, defined in
Eq.~(\ref{eq:amp-dirac}), do not depend on the
Mandelstam variable $t$, and thus the dynamics is governed by the
matrix function $t(s)$ (see Eq.~(\ref{eq:deff0}))
\begin{equation}
t(s) = \sqrt{s}~t_1(s)+t_2(s)  ,
\label{eq:defts}
\end{equation}
which is given in Eq.~(34) of Ref.~\cite{JE01b}. 
There, the
renormalization of the obtained amplitudes is studied at length. 
As a result of the renormalization procedure, and besides the physical
and weak meson decay constants, a total amount of 12
\begin{eqnarray}
J_{\bar K N},&~J_{\pi \Sigma},&~J_{\eta \Lambda},~J_{K \Xi}\nonumber \\ 
 \Delta_N,& ~\Delta_\Sigma,& ~\Delta_\Lambda,~\Delta_\Xi,\nonumber\\ 
  \Delta_{\bar K},&~\Delta_\pi,&~\Delta_\eta,~\Delta_K\label{eq:param}
\end{eqnarray}
undetermined
LEC's appear. We fit these constants to data, as we will see in the
next section, and from them we define the three following diagonal
matrices:
\begin{eqnarray} 
J_0(s=(\hat m + \hat M)^2) &=& \left ( \matrix{ J_{\bar K N}
& 0 & 0 & 0 \cr
0& J_{\pi \Sigma} & 0 & 0 \cr 
0& 0& J_{\eta \Lambda} & 0 \cr
0& 0& 0& J_{K \Xi} } \right ) \nonumber \\ && \nonumber \\ 
 \Delta_{\hat M} &=& \left ( \matrix{  \Delta_{N} 
& 0 & 0 & 0 \cr
0&  \Delta_{\Sigma} & 0 & 0 \cr 
0& 0&  \Delta_{\Lambda} & 0 \cr
0& 0& 0&  \Delta_{\Xi} } \right ) \nonumber \\ 
&& \nonumber \\ 
 \Delta_{\hat m} &=& \left ( \matrix{  \Delta_{\bar K } 
& 0 & 0 & 0 \cr
0&  \Delta_{\pi } & 0 & 0 \cr 
0& 0&  \Delta_{\eta} & 0 \cr
0& 0& 0&  \Delta_{K} } \right ) 
\label{eq:parameters}
\end{eqnarray}
which appears in the solution of the BSE. 
We have denoted the meson-baryon low energy constants $J_0
(s=(m_i+M_j)^2 )$, $i=\bar K,\pi,\eta,K$ and $j=N,\Sigma,\Lambda,\Xi$ of
Eq.~(A8) of Ref.~\cite{JE01b} as $J_{ij}$.  

\section{Numerical results} \label{sec:nr}

Throughout the paper we will use the following numerical values for 
masses and weak decay constants of the pseudoscalar mesons (all in MeV), 
\begin{eqnarray}
& m_K = m_{\bar K}=493.68 \qquad & m_\pi = 139.57 \qquad \ \ m_\eta = 547.3  \nonumber
\\  & M_p = 938.27 \qquad & M_\Sigma = 1189.37  \qquad M_\Lambda =
1115.68 \qquad M_\Xi = 1318.0 \nonumber \\ 
 & f_\pi = f_\eta = f_K = 1.15\times 93.0 
\label{eq:numval}
\end{eqnarray} 
where for the weak meson decay constants we take for all channels an
averaged value\footnote{This makes the comparison with
Ref.~\cite{OR98,ORB02} more straightforward. One could also take the
physical values for these decay constants, but part of the effect can
be absorbed into a redefinition of the parameters in
Eq.~(\ref{eq:parameters}). See also Eq.~(34) in Ref.~\cite{JE01b}.
}. This selection of the coupling constants does not omit the essential features 
of the meson-baryon system since 
a recent work~\cite{Hy02} shows that the observed SU(3) breaking in
meson-baryon scatterings cannot be explained by the present SU(3)
breaking interactions and that the essential physics of the resonances
seems to lie in the substraction constants. 

\subsection{ Fitting procedure} 
\label{sec:data}
We perform a $\chi^2-$fit, with 12 free parameters,  to the following
set of experimental data and conditions:

\begin{itemize}
\item 
\underline{ $S_{01}(S_{2T2J})$ $\bar K N \to \bar K N$ and $\bar K N
  \to \pi \Sigma$ scattering amplitudes 
  (real and imaginary parts)~\cite{Go77}}, \\
\underline{in the CM energy range of  $1480 \, {\rm MeV} \le \sqrt{s} \le
1750 \, {\rm MeV}$}

In this CM energy region, there are a total number of 56 data points
(28 real and 28 imaginary parts) for each channel. The
normalization used in Ref.~\cite{Go77} is different of that used here 
and their amplitudes, $T_{ij}^{{\rm Go77}}$, are related to ours by:
\begin{equation}  
T_{ij}^{{\rm Go77}}={\rm sig}(i,j)|\vec{k}_i|~\left[ f_0^\frac12 (s)
\right]_{j \leftarrow i} ,
\end{equation}
where sig$(i,j)$ is $+1$ for the elastic channel and $-1$ for the $\bar K N
  \to \pi \Sigma$ one.
On the other hand, and because in Ref.~\cite{Go77} errors are not
provided,  we have taken for those amplitudes errors given by 
\begin{equation}
\delta T_{ij}^{{\rm Go77}}=\sqrt{(0.12~T_{ij}^{{\rm Go77}})^2+0.05^2}
\label{eq:Go77}
\end{equation}
in the spirit of those used in Ref.~\cite{Ke01}.

\item 
\underline{ $S_{01}-\pi \Sigma$ mass spectrum 
~\cite{He84}, $1330 \, {\rm MeV} \le \sqrt{s} \le
1440 \, {\rm MeV}$}:
In this CM energy region, there are a total of thirteen 10~MeV bins and 
the experimental data are given in arbitrary units. To compare with
data, taking into account the experimental acceptance
of 10~MeV, we compute:
\begin{eqnarray}
\frac{\Delta\sigma}{\Delta [M_{\pi\Sigma}(i)]}& = &
C~\int_{M_{\pi\Sigma}(i)-5{\rm MeV}}^{M_{\pi\Sigma}(i)
+5{\rm MeV}}
\left|\left [f^\frac{1}{2}_0(s=x^2)\right]_{2\leftarrow 2}\right|^2~|\vec{k}_2(s=x^2)|~x^2~dx \ ,
\end{eqnarray}
where $C$ is an arbitrary global normalization factor\footnote{We fix
 it by setting the area of our theoretical spectrum,
 $\sum_i\frac{\Delta\sigma}{\Delta[ M_{\pi\Sigma}(i)]}$, to
the total number of experimental counts $\sum_i N_i$.} 
and $i$ denotes
the bin with central CM energy $M_{\pi\Sigma}(i)$. Hence, there are only 12 
independent data points. Finally, we take the error of the 
number of counts, $N_i$,  of the bin $i$ to be $1.61\sqrt{N_i}$ as
in Ref.~\cite{Da91}.

\item \underline{The $K^- p \to \eta \Lambda$ total cross section of
    Ref.~\cite{St01}, $1662 \, {\rm MeV} \le \sqrt{s} \le 1684 \, {\rm
      MeV}$}: We use the Crystal Ball Collaboration precise new
total cross-section measurements (a total of 17 data points compiled in
Table I of Ref.~\cite{St01}) for the near-threshold reaction $K^-p \to
\eta\Lambda$, which is dominated by the $\Lambda(1670)$ resonance.
We assume, as in Ref.~\cite{St01}, that the $p-$ and higher wave
contributions do not contribute to the total cross-section. 
\end{itemize} 

Finally, we define the $\chi^2$, which is minimized, as 
\begin{equation}
\chi^2/N_{\rm tot} = \frac{1}{N} \sum_{\alpha=1}^{N}
\frac{1}{n_\alpha}\sum_{j=1}^{n_\alpha}
\left( \frac {x_j^{(\alpha )\,{\it th}}-x_j^{(\alpha
  )}}{\sigma_j^{(\alpha)}}\right)^2\,  ,
\end{equation}
where $N=4$ stands for the four sets of data used and discussed
above\footnote{ From the first item above and to define the $\chi^2$,
  we consider two separated sets: $\bar K N\to \bar K N$ and $\bar K
  N\to \pi \Sigma$.  } 
and $x_j^{{(\alpha )}\,{\it th}}$ denotes
our model result for the data point $x_j^{(\alpha)}$. Finally 
  $n_\alpha$ takes the values $ 56, ~56, ~12$ and $17$, 
and $N_{\rm tot}=\sum_{\alpha=1}^N n_\alpha$ is the total number of
data points. 
With such a definition, acceptable best fits should provide values of
$\chi^2/N_{\rm tot}$ around one.

Though we have considered four coupled channels, 
three-body  channels, for instance
the $\pi\pi\Sigma$ one, are not explicitly considered, as it has been also
assumed previously in Refs.~\cite{ORB02} and~\cite{St01}.

\subsection{ Results of the best $\chi^2$ fit} \label{subsect:best}  
The model presented up to now has initially twelve free parameters
(Eq.~(\ref{eq:param})), which have to be determined from data. This is
a cumbersome task because there are many mathematical minima which are
not physically admissible. For instance, in some cases one finds fits
to data with spurious poles in the first Riemann Sheet which strongly
influence the scattering region and hence violate causality. Any fit
embodying these singularities is physically inadmissible and should be
rejected.  This is an important issue, which should always 
be considered in any analysis. We will further elaborate on this 
point in the Appendix~\ref{app:monster}.

Furthermore, even if one is reasonably convinced that a physically
acceptable minimum has been found, there are strong correlations
between the fitted parameters, which have to be carefully evaluated
and, if possible, understood.  Our best results come from a minimum
for which the pairs $(J_i, \Delta_{B_i})$ with $i = \bar K N, \pi
\Sigma, \eta \Lambda, K \Xi$ and $B_i = N,\Sigma,\Lambda,\Xi$ are
totally correlated (correlation factors bigger than 0.99) which leads
to an almost singular correlation matrix reflecting the fact that
there exists, in very good approximation, a linear relation between
the $J_i$ and $\Delta_{B_i}$ parameters~\footnote{In Ref.~\cite{JE01b}
the static limit (infinitely heavy baryons) is discussed, and it is
shown (Eq.~(D9)) that there exists a linear relation between $J_i$ and
$\Delta_{B_i}$ if, as is the case here, $\Delta_{m_i}$, $i=\bar K,
\pi,\eta,K$ is small when compared to $\Delta_{B_i}$.}.  Thus, we have
fixed $\Delta_{B_i}$ to some specific values in the neighborhood of
the minimum, given in Appendix~\ref{sec:app-stat}, and have studied
the correlation matrix for the remaining eight parameters. Yet, we
find a strong correlation (0.99) between $J_{K\Xi}$ and $\Delta_{K}$
and we proceed as above, i.e., we fix $\Delta_{K}$ and evaluate the
correlation matrix and variances for the remaining seven
parameters. Thus, at the end of the day we have only seven independent
best fit parameters. The best fit parameters, their variances and the
correlation matrix are compiled in Appendix~\ref{sec:app-stat}.

In Figs.~(\ref{fig:1}-\ref{fig:3}) we compare the results of our best
fit with the experimental data. The overall description is remarkably good, both at
low energies and the higher end of the considered energy
region. Besides, as we will see, the description of the
$\Lambda(1405)$ and $\Lambda(1670)$ features is also quite good. 
Thus, our scheme leads to a much better description of the data than the
approach of Ref.~\cite{ORB02}, as it was also the case in the
strangeness $S=0$ sector~(\cite{JE01b} versus \cite{Na00}).

%
%c-------------------------------------------
\begin{figure}[ht]
\vspace{-.75cm}
\begin{center}                                                                
\leavevmode
%\makebox[0cm]{\epsfbox{Rebdivrhor-kfr.ps}}
\makebox[0cm]{
\epsfysize = 150pt
\epsfbox{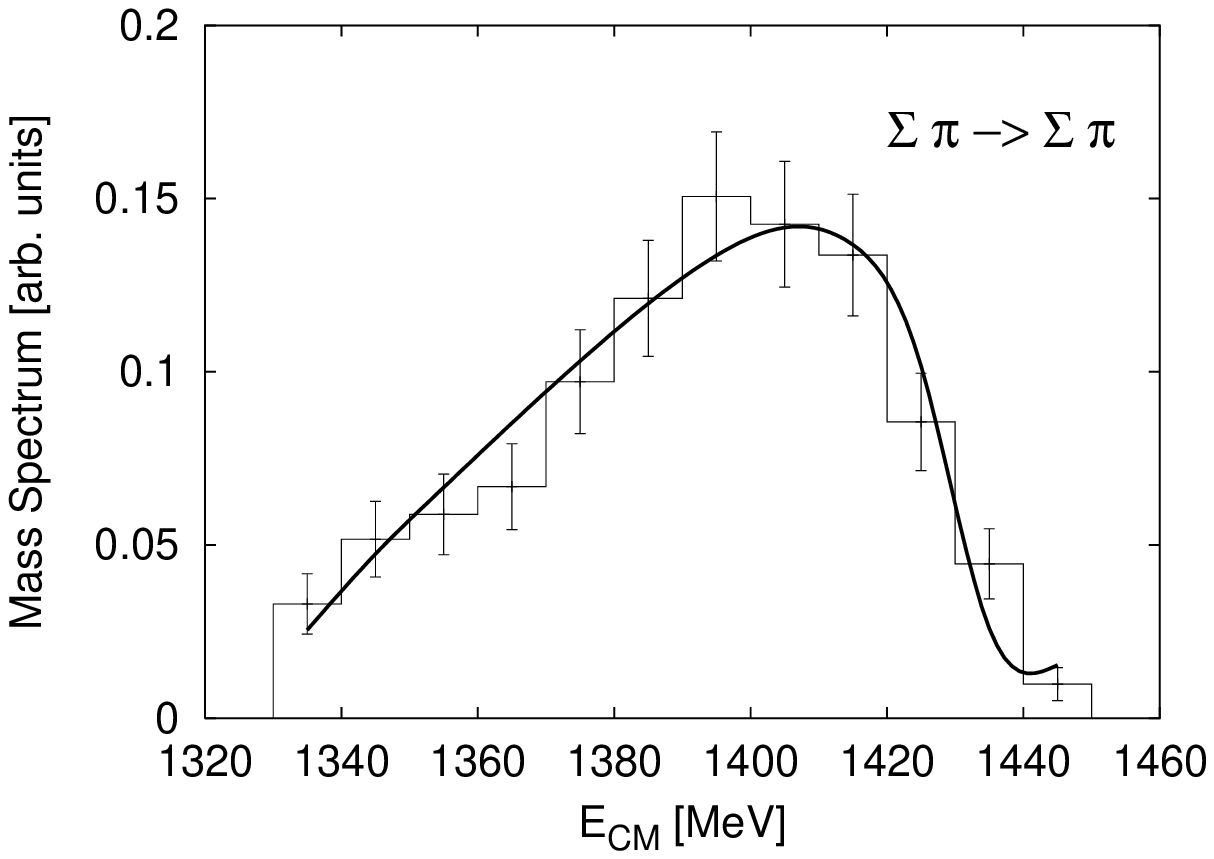}~
\epsfysize = 150pt
\epsfbox{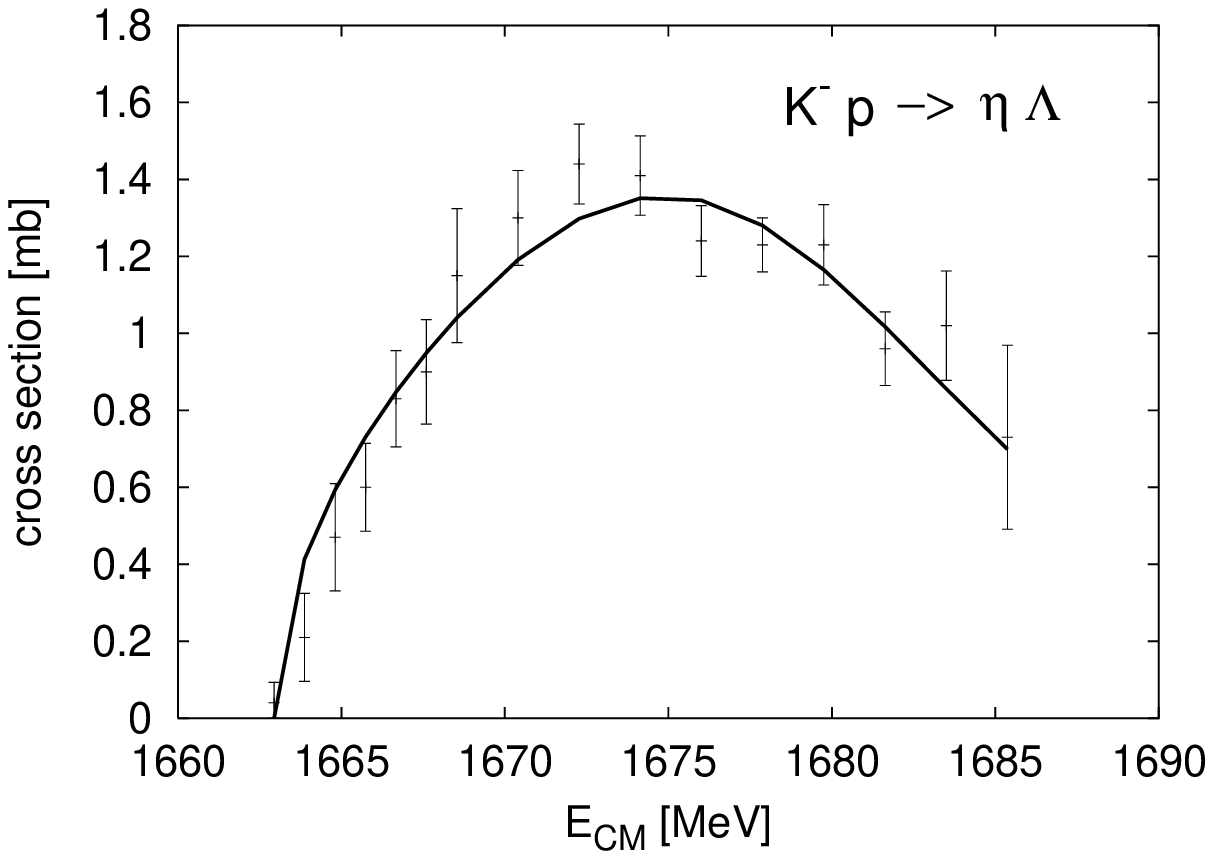}}
\end{center}
\vspace{.5cm}
\caption{{\small Solid lines: Results of our
    fit. Experimental data for $\pi\Sigma\to\pi\Sigma$ and
    $K^- p\to \eta\Lambda$ are from Refs.~\protect\cite{He84}~and~\protect\cite{St01}, respectively }}
\label{fig:1}
\end{figure}
\vspace*{.5cm}
%
%c--------------------------------------------------------------
%c-------------------------------------------
%
\begin{figure}[ht]
\vspace{-.75cm}
\begin{center}                                                                
\leavevmode
\makebox[0cm]{
\epsfysize = 150pt
\epsfbox{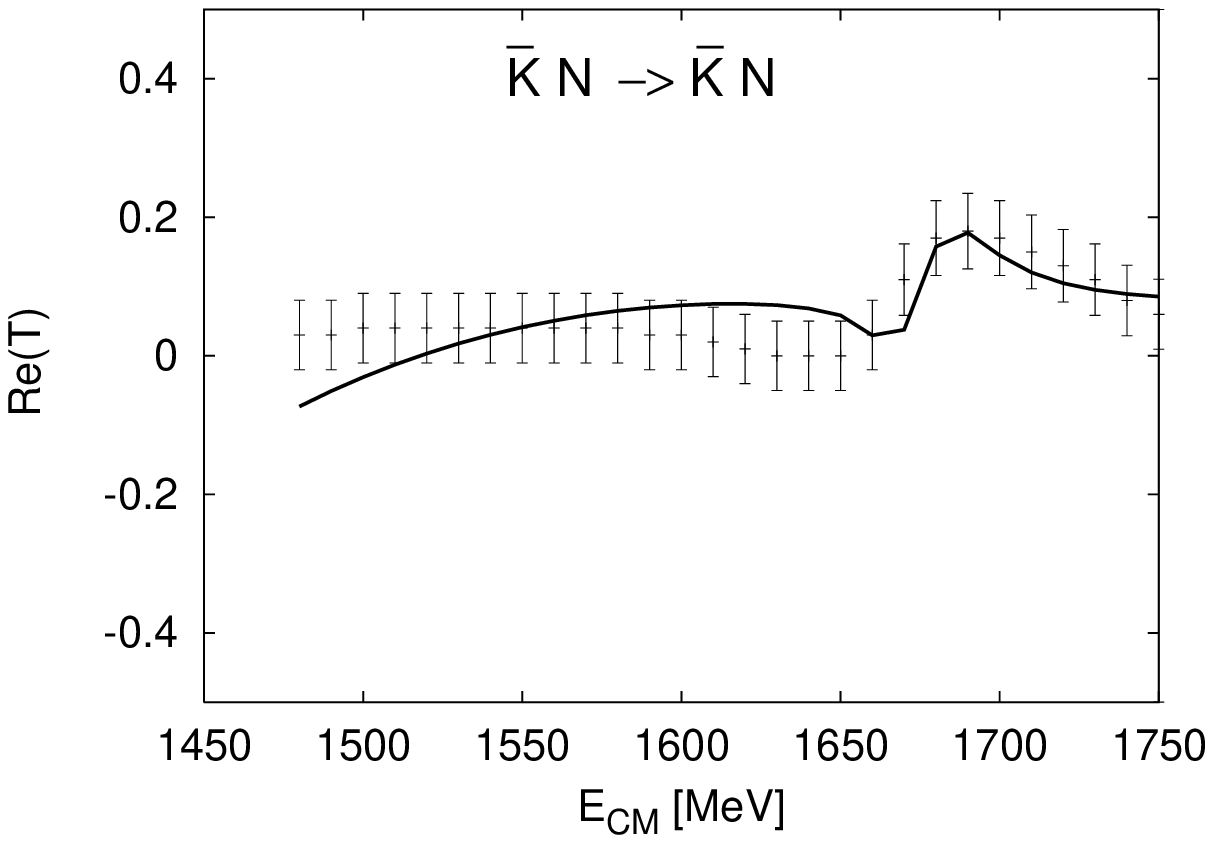}~
\epsfysize = 150pt
\epsfbox{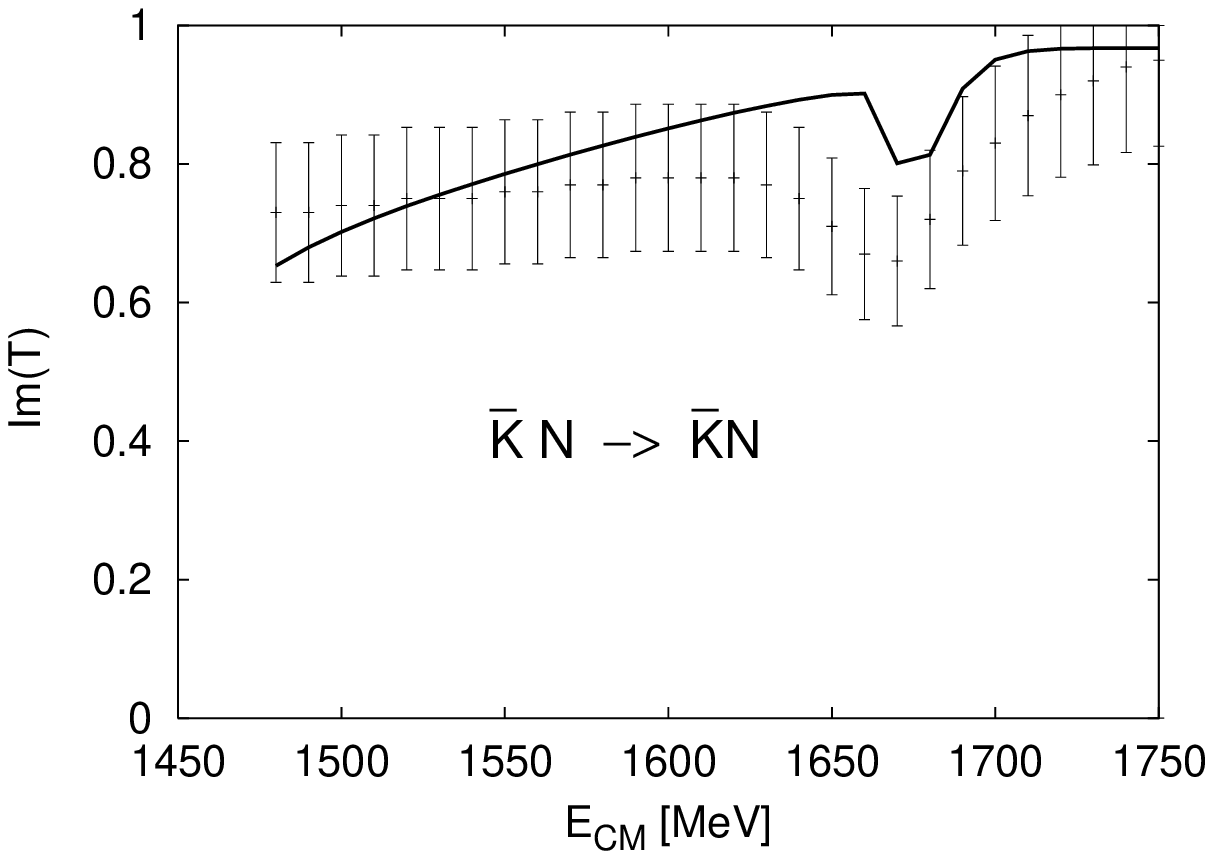}}
\end{center}
%\vspace{-.5cm}
\caption[pepe]{\footnotesize The real (left panel) and imaginary
  (right panel) parts of the  $s-$wave $T-$matrix, with normalization
  specified in Eq.~(\ref{eq:Go77}), 
for elastic $\bar K N \to \bar K N$ process in the $I=0$ isospin channel 
as functions of the CM energy. 
The solid line is the result of our fit, and the experimental data are taken
from the analysis of Ref.~\protect\cite{Go77} with the errors stated in the
main text.}
\label{fig:2}
\end{figure}
\vspace*{.5cm}
%
%c--------------------------------------------------------------
%c-------------------------------------------
%
\begin{figure}[ht]
\vspace{-.75cm}
\begin{center}                                                                
\leavevmode
%\makebox[0cm]{\epsfbox{Rebdivrhor-kfr.ps}}
\makebox[0cm]{
\epsfysize = 150pt
\epsfbox{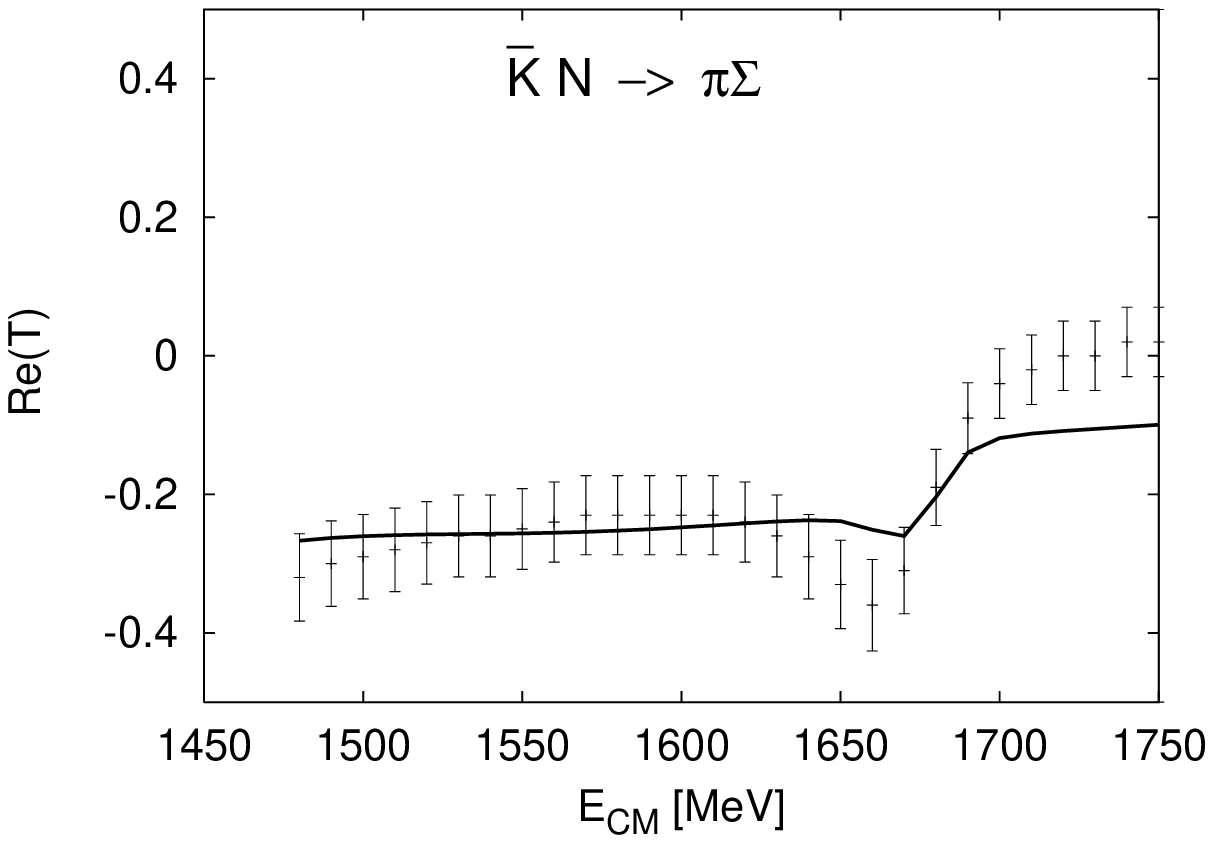}~
\epsfysize = 150pt
\epsfbox{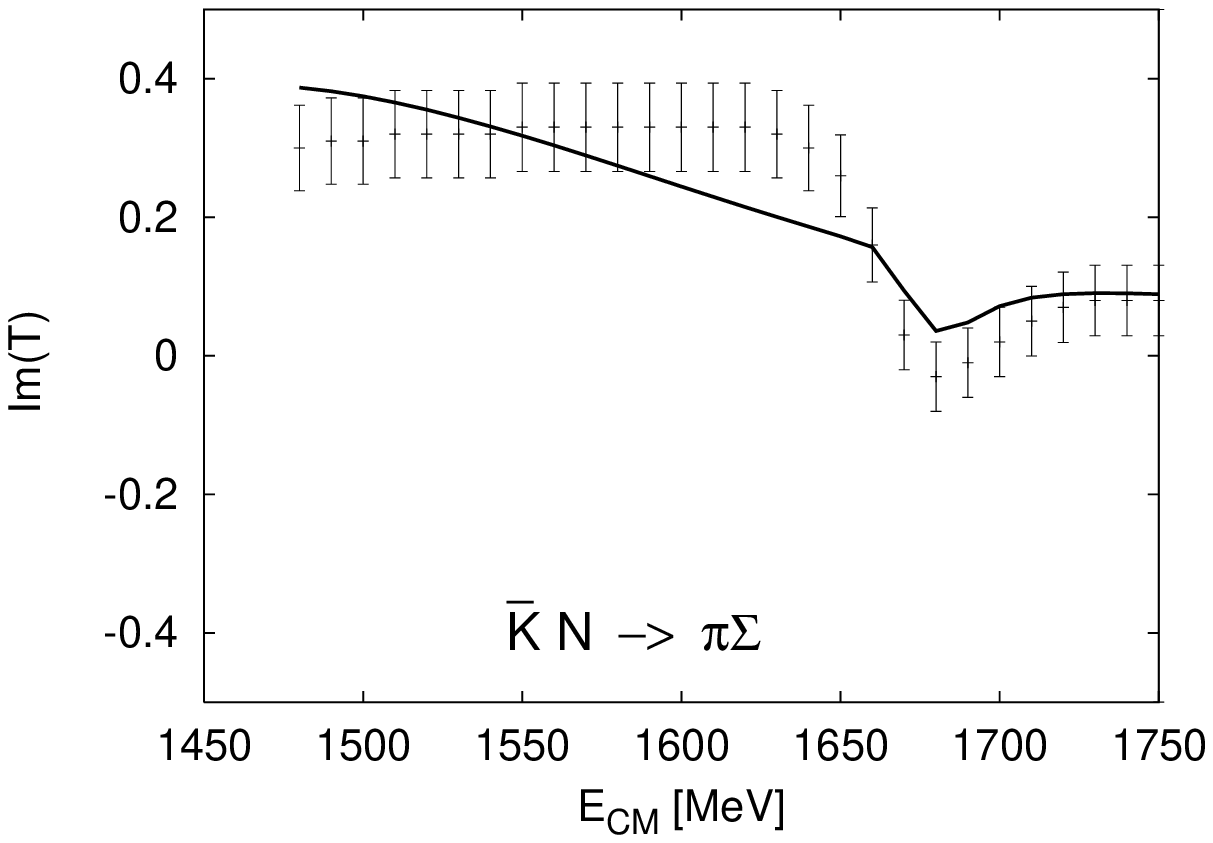}}
\end{center}
\vspace{.5cm}
\caption[pepe]{\footnotesize 
Same as in Fig.~\protect\ref{fig:2} for the 
inelastic channel $\bar K N \to \pi \Sigma$. }
\label{fig:3}
\end{figure}
\vspace*{.5cm}
%
%c--------------------------------------------------------------
For the elastic $\bar K  N \to \bar K N$ scattering length we get
\begin{equation}
a_{\bar K N} \equiv 
\left [ f^\frac12_0(s=(m_K+M_N)^2)\right]_{\bar K N \leftarrow \bar K N} = 
\left (-1.20 \pm 0.09 + {\rm i}~ 1.29 \pm 0.09 \right ) \quad {\rm fm}
\label{eq:apin} 
\end{equation}
where the error is statistical and it has been obtained from the
covariance matrix given in the Appendix~\ref{sec:app-stat}, taking
into account the existing statistical correlations, through a
Monte--Carlo simulation. This value should be compared both to the
experimental one $(-1.71 + {\rm i}~ 0.68) $~fm of Ref.~\cite{Ma81} and
to the LSE approach of Ref.~\cite{OR98} $(-2.24 + {\rm i}~ 1.94)
$~fm. Unfortunately, the previous works do not provide error
estimates, so one cannot decide on the compatibility of results.

\subsection{ Second Riemann Sheet: poles and resonances.} 
\label{sec:reso}

In this section we are interested in describing masses and widths of
the $S_{01}-$ resonances in the $S=-1$ channel.  Since causality
imposes the absence of poles in the $t(s)$ matrix in the physical
Sheet~\cite{Ma58}, one should search for complex poles in unphysical
ones.  Among all of them, those {\it closest} to the physical Sheet
and hence to the scattering line are the most relevant ones. We define
the second Riemann Sheet in the relevant fourth quadrant as that which
is obtained by continuity across each of the four unitarity cuts (see
a detailed discussion in a similar context in Ref.~\cite{JE01b}).
Physical resonances appear in the second Riemann Sheet of all 
matrix elements of $t(s)$, defined in Eq.~(\ref{eq:defts}), in
the coupled channel space, differing only on the value of the residue
at the pole. The residue determines the coupling of the resonances to the
given channel. In Fig.~\ref{fig:riemman} we show the absolute value of the
$\eta\Lambda \to \eta\Lambda$ element of the $t$ matrix.  We choose
this channel because all found poles have a sizeable coupling to it.
Both the fourth quadrant of 
the second Riemann Sheet and the first quadrant of the first
(physical) Riemann Sheet are shown. The physical scattering takes place in the
scattering line in the plot (upper lip of unitarity cut of the first
Riemann Sheet). We find three poles in  the second Riemann
Sheet which positions  are ($s = M_R^2 - {\rm i} M_R \Gamma_R$):
\vspace{-0.2cm}
\begin{figure}
\epsfysize = 350pt
%\epsfbox{h33-7-arrow.ps}
\epsfbox{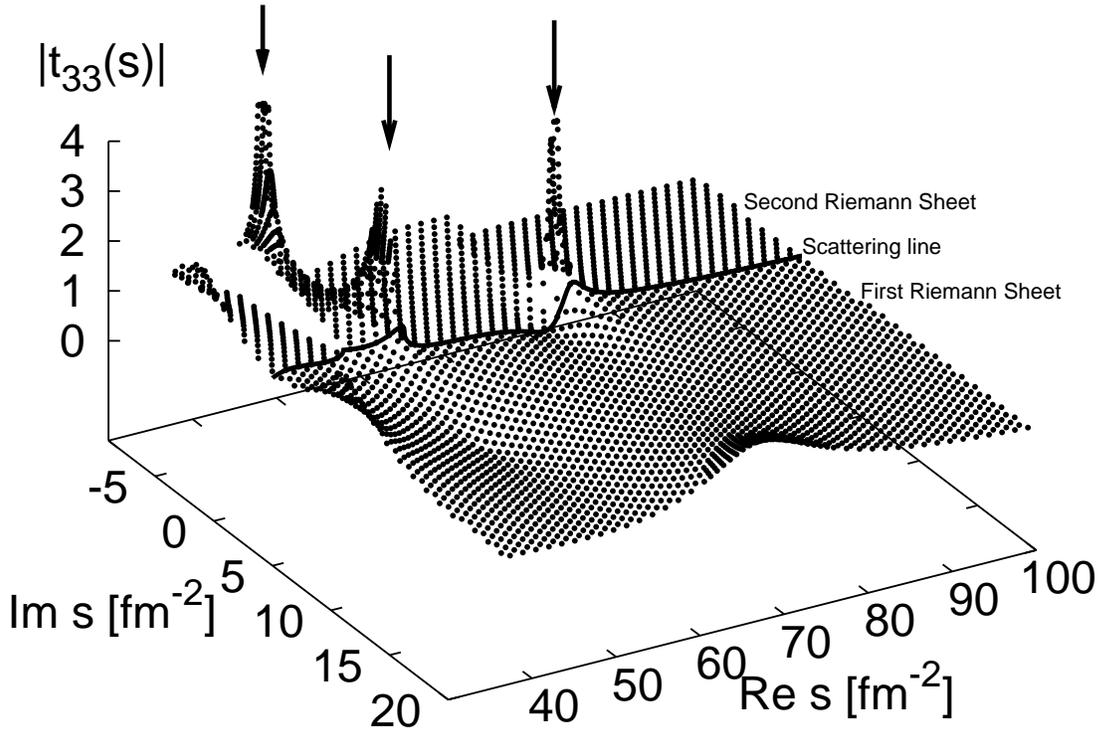}
%\centerline{\epsfig{figure=h33-7-3.ps,height=12cm,width=17cm}}
\caption{ \footnotesize Modulus of the $\eta \Lambda \to \eta \Lambda$
element of the scattering amplitude $t(s)~[\rm{fm}]$, defined in
Eq.~(\protect\ref{eq:defts}), analytically extended to the first and
fourth quadrants of the $s-$complex plane. The solid line is the
scattering line, $s= x+{\rm i}~0^+,~x\in \protect\reales$, from the first
threshold, $(m_\pi+M_\Sigma)^2$, on. The first (second) Riemann Sheet
is depicted in the first (fourth) quadrant of the $s$ complex plane.
Three poles appear in the second Riemann Sheet, which are connected
with the $\Lambda(1405)$ and $\Lambda(1670)$ resonances, see
discussion in the main text. Besides unphysical poles show up in the
physical Sheet out of the real axis, but they do not influence the
scattering line as can be seen in the plot. }
\label{fig:riemman}
\end{figure}
\begin{eqnarray}
{\rm First~~~ Pole:~~}  M_R &= 1368\phantom{.5} \pm 12 \qquad  &\Gamma_R
= 250 \pm 23 \label{eq:1300}\\
&&\nonumber\\  
{\rm Second~~~ Pole:~~}  M_R &= 1443\phantom{.5} \pm \phantom{5}3 \qquad  &\Gamma_R
= 50 \pm 7 \label{eq:1400}\\
&&\nonumber\\  
{\rm Third~~~ Pole:~~}  M_R &= 1677.5 \pm 0.8 \qquad  &\Gamma_R = 29.2 \pm
1.4  \label{eq:1670}
\end{eqnarray}
where all units are given in MeV and errors have been transported from
those in the best fit parameters (Eq.~(\ref{eq:lecs1})), taking into
account the existing statistical correlations through a Monte--Carlo
simulation.

These poles are related to the two $S_{01}$
resonances $\Lambda(1405)$ and $\Lambda(1670)$ which appear up to this
range of energy in the PDG (Ref.~\cite{pdg02}). The third pole above can be
clearly identified to the $\Lambda(1670)$ which  is located at  
\begin{eqnarray}
 \Lambda(1670):~~~ & M_R = 1670\pm 10 \qquad  \Gamma_R
= 35 ^{+15}_{-10} & {\qquad\rm{Ref}.~\protect\cite{pdg02}}\nonumber \\
\phantom{\Lambda(1670):}~~~ & M_R = 1673\pm\phantom{0}2 \qquad  \Gamma_R
= 23 \pm 6 & {\qquad\rm{Ref}.~\protect\cite{Ma02}}
\end{eqnarray}
where again units are in MeV.  The agreement of our predictions and
the experimental data is satisfactory and better than the previous
theoretical LSE approach of~Ref.~\cite{ORB02}.
Let us look at the $\Lambda(1405)$ resonance, 
which nature is under much discussion~\cite{Dalitz:jx,Kimura:sm}. 
Following the  PDG it is
placed at (in MeV) 
\begin{eqnarray}
 \Lambda(1405):~~~ & M_R = 1406.5\pm 4.0 \qquad  \Gamma_R
= 50 \pm 2 & {\qquad\rm{Ref}.~\protect\cite{pdg02}}  
\end{eqnarray}
 Our amplitudes have two poles in the region of
1400~MeV, Eqs.~(\ref{eq:1300}) and~(\ref{eq:1400}). The features of
the second one are in agreement with the previous results of 
Refs.~\cite{OR98,ORB02} and though the width compares well with the
experiment, the mass is shifted to higher values. Besides, we should
note that the pole quoted in Eq.~(\ref{eq:1300}) is very broad and can
not be identified with any of the experimentally established
resonances. This pole is also present in the LSE model of
Refs.~\cite{OR98,ORB02}, as it was pointed out in Ref.~\cite{Ji01},
though the mass position there is similar ($M_R=1390$~MeV), the width is about
a factor two narrower ($\Gamma_R=132$~MeV) than ours. 
Our understanding is that this broad resonance does not influence strongly the
scattering line. However, the $\pi \Sigma$ mass spectrum peaks around
1405 MeV  in the  experimental data and also in our approach as can be
seen in Fig.~\ref{fig:1}. This is a clear indication of a sizeable 
non resonant contribution on top of our 1443~MeV pole. 

On the other hand, there are unphysical poles in the physical (first)
Riemann Sheet. These unphysical poles appear because we have truncated
the iterated potential to solve the BSE.  The two of them closer to
the scattering line are located at ($s = M^2 + {\rm i} M \Gamma$) with
$M \approx 1166,~\Gamma\approx \pm 200$~MeV and $M \approx
1616,~\Gamma\approx 631$~MeV. The tails of both poles can be seen in
Fig.~\ref{fig:riemman} and they do not influence the scattering line.
In Appendix~\ref{app:monster}, we will show the results from a fit
which, at first sight, are even in a better agreement with the
experimental data (Subsect.~\ref{sec:data}) than those presented up to
now.  However this apparent improvement is achieved because the
unphysical poles get closer to the real $s$ axis and they affect, in a
substantial manner, the scattering amplitudes. Hence, we discard this
minimum, and we would like to note that it is important to observe the
positions and influence of the unphysical poles when deciding the
goodness of a phenomenological description of data. 

Finally, we have also analyzed the nature of resonances on the light
of the well known Breit-Wigner parameterization for coupled channels
and real $s$
(See e.g. Ref.~\cite{BK82} and references therein), 
\begin{eqnarray} 
t_{ij}^{\rm BW} (s) = -\frac{\delta_{ij}}{2 {\rm i} \rho_i } \left[
e^{2 {\rm i} \delta_i } - 1 \right] + \frac{e^{{\rm i} ( \delta_i +
\delta_j ) } M_R \sqrt{ \Gamma_i^{\rm BW} \Gamma_j^{\rm BW}} }{
\sqrt{\rho_i \rho_j } \left[ s - M_R^2 + {\rm i} M_R \Gamma_R \right]
}
\label{eq:bw} 
\end{eqnarray} 
for which the background is assumed to be diagonal in coupled channel
space and the relative phase of the resonance to the background and
the sum partial decay widths, $\sum_i \Gamma_i^{\rm BW} = \Gamma_R $,
are chosen in such a way that $t_{ij}^{\rm BW} (s)$ exactly fulfills
unitarity on the real axis. Here, $\rho_i$ is a kinematic factor defined
by the second line of Eq.~(\ref{eq:sres}) below. 
The branching ratio is then defined as
$B_i^{\rm BW}= \Gamma_i^{\rm BW} / \Gamma_R $. Subtracting the
resonance contribution, Eq.~(\ref{eq:tpole}), to the total amplitude 
we have found that for 
our $\Lambda (1670) $ the background is not a diagonal matrix, since
for our $t_{ij}$-matrix we get $ 2 \sum_{i<j} |t_{ij} - t_{ij}^{\rm
RBW} |^2 \approx \sum_{i} |t_{ii} - t_{ii}^{\rm RBW} |^2 $ for $ s \to
M_R^2 $, with $ t_{ij}^{\rm RBW} $ the second term in
Eq.~(\ref{eq:bw}). In addition, the BW parameterization suggests a
relation between the residue at the pole and the imaginary part of the
pole. This relation is only true in the sharp resonance approximation,
$ \Gamma_i^{\rm BW} << p_i $ with $p_i$ the CM momentum of the
decaying state. We have also checked that for our problem this is not
the case.  Actually, with such a definition we find that $\sum_i
\Gamma_i^{\rm BW} \approx 0.8 \Gamma_R $ for the $\Lambda(1670)$. This
is a simple consequence of the incorrect assumption made by
Eq.~(\ref{eq:bw}).

\subsection{Branching ratios and couplings of the resonances to
different final states.}
Before going further we would like to make some critical remarks
regarding both the comparison between ``theory'' and
``experiment''. Our BSE solution has a very specific energy dependence
which, as we have seen in Subsect.~\ref{subsect:best}, is able to fit
numerically experimental data or rather a partial wave analysis with a
given energy dependence. Obviously, both functional forms are not
identical, and it is also fair to say that both incorporate their own
biases. There is no reason to expect that they are numerically alike
also in the complex plane~\footnote{A good example of this fact is
provided by our best fit results of Subsect.~\ref{subsect:best} and
the physically inadmissible results of Appendix~\ref{app:monster}; they
look  very much the same on the scattering line although the
analytic structure is rather different.}. Under these conditions, some
parameters, like branching ratios, have a different meaning, since the
extrapolation of the resonant contribution to the real s-axis is
ambiguous. Actually, the ambiguity is enhanced as the resonance
becomes wider and as a consequence the definition of a branching ratio
becomes model dependent. We explain below our definition of branching
ratios and how they are extracted from our amplitude. 

Let us consider $s_R = M^2_R - {\rm i}~ M_R \Gamma_R$ a pole in the
second Riemann Sheet of the coupled channel scattering matrix
$t(s)$. Then, around the pole, it can be approximated by
\begin{eqnarray}
\left[t(s)\right]_{ij} & \approx & 2 M_R \frac{g_{ij}}{s-s_R},
\label{eq:tpole}
\end{eqnarray}
where $g_{ij}$ is the residue matrix. Since $t$ is a complex symmetric
matrix (due to time reversal invariance), $g$ is also complex
symmetric and its rank is one to ensure that $s=s_R$ is a pole of
order one of the ${\rm det}[t(s)]$. In this way a non-degenerate
resonant state is being described\footnote{This can be seen as
follows. Using a matrix notation, the BSE reads $ t(s)= V + V G_0 (s)
t(s) $ and it is solved by $t(s) = V (1-G_0(s)V)^{-1}$, with the
obvious identifications for $V$ and $G_0$. A pole, at $s=s_R$, in
${\rm det}[ t(s)] $ is produced by a zero of ${\rm det} [1-G_0(s)V
]$. This last condition ensures that  the homogeneous
(quasi)bound state Bethe-Salpeter equation
\begin{equation} 
 (-G_0(s)^{-1} +V)\Psi = 0 
\nonumber
\end{equation}
has a non trivial solution for $s=s_R$. Indeed, all solutions, $\Psi$,
of the above equation are linear combination of the null eigenvectors
of the $(1-G_0(s_R)V)$ matrix and describe the dynamics of the
existing states (resonances) at $s=s_R$. For a non-degenerate
resonance (which we have checked it is indeed the case), the zero
eigenspace should have dimension one. Whence for the case of four
coupled channels, the rank of the matrix $(1-G_0(s_R)V)$ is three and
therefore ${\rm det}[1-G_0(s)V]$ should have a single zero at
$s=s_R$. }. Under these conditions, $g_{ij}$ turns out to be
factorizable\footnote{The symmetric complex matrix $g$ can be
diagonalized by a complex orthogonal transformation, $U$, as $g=U^T~ d
~ U$, being only one element of the diagonal matrix $d$ different of
zero. If we take this element to be the $d_{11}$, we have $g_{ij}=
(U^T)_{i1}d_{11} U_{1j} = d_{11}U_{1i} U_{1j}$ .},
\begin{equation}
 g_{ij}= g_i g_j  .
\label{eq:gigj}
\end{equation}
The above matrix, $g_{ij}$, has only one non-zero eigenvalue,
$g_1^2+g_2^2+g_3^2+g_4^2$, with $g_i$ the associated eigenvector.  The
vector $g_i$ determines the coupling of the resonance to the different
final states, which are well and unambiguously defined even if the
corresponding channels are closed in the decay of the resonance. In
Table~\ref{tab:gi} we give the complex vectors $g_i$ for the three
resonances described in Subsect.~\ref{sec:reso}. Unfortunately, the
PDG does not provide this kind of information and, instead, branching
ratios are given. To extract meaningful branching ratios from our
calculation, we have to extrapolate the resonant contribution of the
scattering amplitude to the $s$ real axis, which is the only
experimentally accessible. In addition, the picture of a resonance as
a quantum mechanical decaying state, requires a probabilistic
description.  Thus, we isolate the resonant contribution to the
$S-$matrix\footnote{The $S-$matrix is related to the $t$ matrix, in
our convention, by:
\begin{equation}
S_{ij}(s) = \delta_{ij}-2{\rm i} ~\sqrt{\rho_i(s)}~t_{ij}(s)~\sqrt{\rho_j(s)}
\end{equation}
and probability conservation ($S^\dagger S = SS^\dagger = 1$) holds
since $t$ fulfills coupled channel unitarity:
\begin{equation}
t^*_{ij}(s) - t_{ij}(s) = 2{\rm i}~\sum_k t^*_{ik}(s)~ \rho_k(s)~
t_{kj}(s) .
\end{equation}
} for $s=M^2_R$
\begin{eqnarray}
S^{\rm resonant}_{ij}(s=M^2_R) &=& - 2 {\rm i}~ 2
M_R \sqrt{\rho_i^R}  \frac{g_ig_j}{M^2_R-s_R}
\sqrt{\rho_j^R} \nonumber \\
\rho_i(s) &=&  \Theta(s-s_{th}^i) \frac{|\vec{k}_i(s)|}{8\pi\sqrt{s}} 
\left (\sqrt{M_i^2+\vec{k}_i^2}+M_i\right)
\nonumber\\
\label{eq:sres}
\rho_i^R &=& \rho_i(s=M_R^2)
\end{eqnarray}
with $s_{th}^i$ the threshold of the baryon-meson channel $i$. This
definition embodies a sensible kinematic suppression compatible with
Cutkosky's rules and the s-wave nature of the resonance. Defining
\begin{equation}
b_i = g_i \sqrt{\frac{2 \rho_i^R}{ \Gamma_R}}  \label{eq:bi}
\end{equation}
we find
\begin{eqnarray}
S^{\rm resonant}_{ij}(s=M^2_R) &=& - 2 {\rm i}~ 
M_R \Gamma_R \frac{b_ib_j}{M^2_R-s_R}
\end{eqnarray}
and taking into account that the matrix $b_{ij} = b_i b_j$ has rank 1
and that $b_i$ (or any vector proportional to it)  is the only
eigenvector of $S^{\rm resonant}$ with a non zero eigenvalue, the resonant state at 
$s=M^2_R$ will be given by 
\begin{equation}
|R \rangle \propto \sum_i  b_i |i \rangle
\end{equation}
where $|i \rangle $ stands for the meson-baryon states used to
build the coupled channel space. Finally, the branching ratio $B_i$ 
will be given by the probability of finding  $|R\rangle$ in the state $|i\rangle$
\begin{equation}
B_i = \frac{|b_i|^2}{\sum_j |b_j|^2}
\end{equation}
which by definition fulfills $\sum_i B_i = 1$. The partial decay
width may then be defined as $\Gamma_i = B_i \Gamma_R $, and obviously
$ \sum_i \Gamma_i = \Gamma_R $. For the $\Lambda(1670)$ we obtain the
following branching ratios with the above prescription:
\begin{equation}
B_{\bar K N} = 0.24\pm 0.01, \qquad B_{ \pi \Sigma} = 0.08\pm 0.01,
\qquad B_{ \eta \Lambda} = 0.68\pm 0.01   \qquad 
\end{equation}
The last two values are not in agreement with the values 
quoted in the PDG~\cite{pdg02}  
(
$B_{\bar K N}= 0.25\pm 0.05, ~ B_{ \pi \Sigma}= 0.40\pm 0.15,~ B_{
 \eta \Lambda}= 0.17\pm 0.07$
)
and in Ref.~\cite{Ma02}
(
$B_{\bar K N}= 0.37\pm 0.07, ~ B_{ \pi \Sigma}= 0.39\pm 0.08,~ B_{
 \eta \Lambda}= 0.16\pm 0.06,~B_{\pi\Sigma(1385)}=0.08\pm 0.06$
). 
{
\begin{table}
\begin{center}
\begin{tabular}{l|cc|cc|cc|cc}\hline\tstrut 
%\hline\tstrut
 & \multicolumn{2}{c|}{$g_{\bar K N}$}&\multicolumn{2}{c|}
{$g_{\pi\Sigma}$ }&\multicolumn{2}{c|}{  $g_{\eta\Lambda}$} &
\multicolumn{2}{c}{$g_{K \Xi }$ }\\
Resonance  & $|g|$ & $\phi$ & $|g|$ & $\phi$ & $|g|$ & $\phi$ & $|g|$ & $\phi$ \\\hline
$M_R=1368$~MeV & 3.9(1) & -0.59(5)& 3.65(8) & -0.73(3) &
 1.7(2) & 3.0(2) &0.29(7) & 1.14(13) \\
$M_R=1443$~MeV & 3.3(2) & 0.72(7) & 2.14(16) & 1.10(8) & 
2.2(1) & -2.66(3) & 0.23(1) & -0.008(57)\\
%2.2(1) & -2.66(3) & 0.23(1) & -0.008 !!!(53)!!! \\ !!! cambio respecto
%enviado PRD el 18-Oct-2002
$M_R=1677.5$~MeV & 0.39(2) & -1.29(4) &  0.20(1) & 0.77(5)& 
 1.22(3) & 2.69(2) & 1.64(1) & -0.13(1)
 \end{tabular}
 \end{center}
 \caption{ {\protect\small Dimensionless complex couplings, $g_i=|g_i|~e^{{\rm
       i}\phi_i}$, defined in Eqs.~(\ref{eq:tpole}-\ref{eq:gigj}), for all channels $i=
 \bar K N,~\pi\Sigma,~\eta\Lambda,~K \Xi$, and for the three poles
 (resonances) quoted in Eqs.~(\protect\ref{eq:1300}--\protect\ref{eq:1670}). 
The phases are in radian units. Errors are purely statistical and 
affect the last significant digit.}}
 \label{tab:gi}
 \end{table}}

To finish this subsection, we would like to point out that in the
present context the concept of branching ratio is subtle and it might
be ambiguous, both from the theoretical and experimental sides.  From
the experimental point of view the difficulty arises from the
impossibility of preparing a pure short-lived resonant state strongly
coupled to a continuum and therefore the impossibility of disentangle
events coming from the formation of the resonance from those produced
through non-resonant processes. From the theoretical point of view the
ambiguity comes when defining $S^{\rm resonant} (s= M^2_R)$ in
Eq.~(\ref{eq:sres}).  For instance at the pole $s=s_R$ one could have
\begin{eqnarray}
\left[t(s)\right]_{ij} & \approx & 2 M_R
\frac{\beta_i(s)}{\beta_i(s_R)}~\frac{g_{ij}}{s-s_R}~
\frac{\beta_j(s)}{\beta_j(s_R)}, 
\end{eqnarray}
instead of the expression assumed in Eq.~(\ref{eq:tpole}), being
$\beta_i(s)$ an arbitrary, analytical around $s_R$, complex
function. In these circumstances the matrix $S^{\rm resonant}_{ij} (s=
M^2_R)$ would be different from that given in Eq.~(\ref{eq:sres}) by a
factor $ \frac{\beta_i(M^2_R)}{\beta_i(s_R)}
\frac{\beta_j(M^2_R)}{\beta_j(s_R)} $  and one would get a new vector
$\tilde{b}_i$ which in terms of the vector $b_i$, defined in
Eq.~(\ref{eq:bi}), reads:
\begin{eqnarray}
\tilde{b}_i =  \frac{\beta_i(M^2_R)}{\beta_i(s_R)}~ b_i
\end{eqnarray}
leading to, in principle, different branching ratios. 
The trouble comes from the extrapolation from $s_R$ to the
real axis which is not unique.

The usual assumption is that the $\beta_i(s)$ functions are smooth and
they do not change much from $s=s_R$ to $s=M_R^2$, and more important,
that the change does not depend significantly on the channel $i$.
However, this is not always true, for example, if one were dealing
with a $p-$wave resonance the function $\beta_i(s)$ would at least
include the CM meson baryon momentum.  If there is a channel which
gets open close and above to $s=M_R^2$, then the CM momentum would
lead to a suppression of the branching ratio to this channel.

\subsection{ Predictions for other processes} 
\begin{figure}
\begin{center}                                                               
\leavevmode
\epsfysize=250pt
\makebox[0cm]{\epsfbox{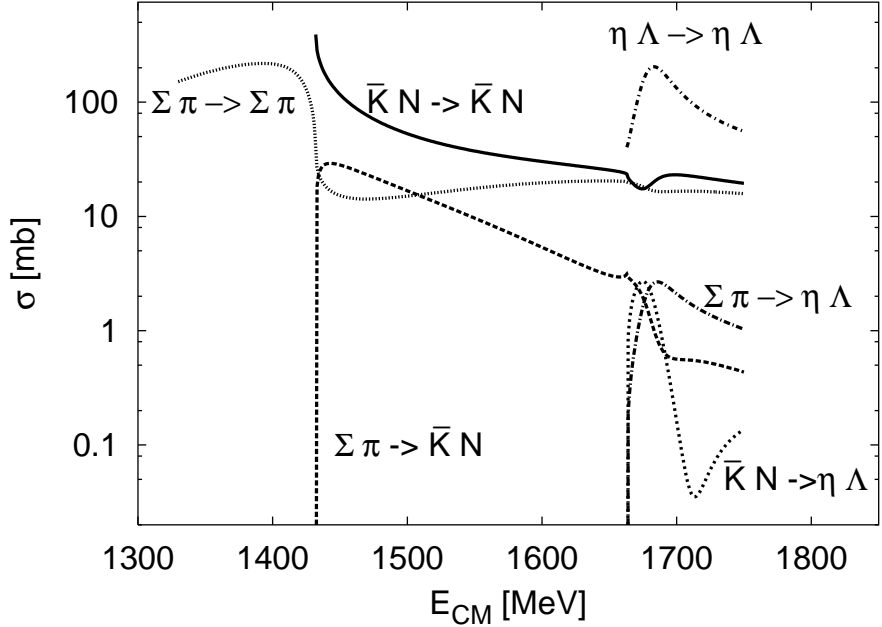}}
\end{center}
%\vspace{-3.4cm}
\caption[pepe]{\footnotesize 
$I=0$  meson-baryon $s-$wave cross
sections for different channels.}
\label{fig:sigma-t}
\end{figure}

In Fig.~\ref{fig:sigma-t} we show some of our predictions for $s-$wave
 $I=0$ cross sections for some elastic and inelastic channels. For
 most of them there are no data.  The effect of the $\Lambda(1405)$
 resonance is clearly visible in the $\Sigma \pi \to\Sigma \pi $ and
 $\Sigma \pi \to\bar K N $ and $\bar K N \to\bar K N$ cross sections.
 On the other hand, the elastic $\eta \Lambda$ cross section takes a
 very large value at threshold, which corresponds to a typical low
 energy resonance behavior triggered by the $\Lambda(1670)$
 resonance. This is in contrast to any expectation based in the Born
 approximation, since the corresponding potential in this channel
 vanishes~(Eq.~(\ref{eq:d-matrix})).
%
%\vspace{-2cm}
Our estimates for $\pi \Sigma$ and $\eta \Lambda$ scattering lengths (defined
for the elastic channels similarly as in Eq.~(\ref{eq:apin})) are
\begin{eqnarray} 
a_{\pi \Sigma}    &=& 1.10\pm 0.06 \, {\rm fm}\nonumber  \\ 
a_{\eta \Lambda} &=& (0.50\pm0.05 )\, + \, {\rm i} \, ( 0.27\pm0.01) \, {\rm fm} 
\end{eqnarray} 
respectively. 

\subsection{Heavy Baryon Expansion at Threshold} 
As we have already mentioned, the only calculation within HBChPT in the
$S=-1$ sector is that of Ref.~\cite{Ka01}, where the $s$-wave
scattering lengths in both isospin channels $I=0$ and $I=1$ are
computed. It is found that HBChPT to one loop fails completely in the
$I=0$ channel due to the strong influence of the subthreshold
$\Lambda(1450)$ resonance. We think that it is of interest to analyze this
problem within the context of our unitarization approach. 

As was discussed previously by two of us~\cite{JE01b} the condition
for the coupled channel amplitude to have a well defined static limit,
$M \to \infty $, is that the combination
\begin{eqnarray} 
C_{\hat m \hat M} = \frac1{\hat m} \left[ \hat M J_{\hat m\hat M} +
\frac1{4 \hat M} (\Delta_{\hat m} -\Delta_{\hat M}) \right]
\end{eqnarray}  
goes to some definite finite value.  The parameters $J_{\hat m\hat M}$
and $\Delta_{\hat m}$ and $\Delta_{\hat M}$ are listed in
Eq.~(\ref{eq:parameters}) and their best fit result is presented in
Eq.~(\ref{eq:lecs1}) of Appendix~\ref{sec:app-stat}.  
Using Eq.~(\ref{eq:lecs1}) one can
estimate the combinations in units of the relevant meson mass
\begin{eqnarray} 
M_{\Sigma} J_{\pi \Sigma}+ \frac{\Delta_{\pi}-\Delta_{\Sigma}}{4
M_{\Sigma} } &=&  +0.24 m_\pi \nonumber \\ M_N J_{\bar K N}+
\frac{\Delta_{\bar K}- \Delta_N}{4 M_N } &=& -0.09 m_{\bar K}
\nonumber \\ M_\Lambda J_{\eta \Lambda}+
\frac{\Delta_{\eta}-\Delta_{\Lambda}}{4 M_\Lambda} &=& +0.08 m_{\eta}
\nonumber \\ M_\Xi J_{K \Xi   }+ \frac{\Delta_{K}-\Delta_\Xi}{4
M_\Xi} &=& -0.64 m_K 
\end{eqnarray} 
showing that they are not unnaturally large. Following the discussion
of Ref.~\cite{JE01b}, the heavy baryon expansion is done in the
standard way, and taking the leading heavy baryon approximation of the
parameters,  
\begin{eqnarray} 
J_{\hat m \hat M} &=& J_{\hat m \hat M}^0 \left\{ 1 + {\cal O} \left(
\frac1M \right)\right\} \nonumber \\
\Delta_{\hat m} &=& \Delta_{\hat m}^0 \left\{ 1 + {\cal O} \left(
\frac1M \right)\right\}  \\ \Delta_{\hat M} &=& 
\Delta_{\hat M}^0 \left\{ 1 + {\cal O} \left( \frac1M \right)\right\} \nonumber
\label{eq:hbexp}
\end{eqnarray} 
We get for the $\bar K N$ s-wave scattering length in the $I=0$
channel the following expressions
\begin{eqnarray}
a_{\bar K N} &=&\frac{3 m_{\bar K} }{8 \pi f_K^2}\left[ 1-
\frac{m_{\bar K}}{M_N} + \frac{m_{\bar K}^2}{M_N^2} \right] +
\frac9{16 \pi f_K^4} \left( \frac34 \Delta_{\bar K}^0 m_{\bar K} - 2
C_{\bar K N}^0 m_{\bar K}^3 \right) \nonumber \\ &+& \frac3{32 \pi
f_K^2 f_\pi^2} \left\{ \frac34 \Delta_\pi^0 m_{\bar K} - 2 C_{\pi
\Sigma}^0 m_{\bar K}^2 m_\pi - \frac1{ 4\pi^2 } m_{\bar K}^2
\sqrt{m_{\bar K}^2-m_\pi^2} \left[
{\rm~arccosh}\left(\frac{m_\pi}{m_{\bar K}} \right) - {\rm i} \pi
\right] \right.  \nonumber \\ && \qquad \qquad - \left. \frac1{4\pi^2}  m_{\bar K}^2 
(m_\pi-m_{\bar K}) \ln \frac{M_\Sigma}{m_\pi} \, \right\} \nonumber \\
&+& \frac9{32 \pi f_K^2 f_\eta^2} \left\{ \frac34 \Delta_\eta^0
m_{\bar K} - 2 C_{\eta \Lambda}^0 m_{\bar K}^2 m_\eta - \frac1{ 4\pi^2
} m_{\bar K}^2 \sqrt{m_{\eta}^2-m_{\bar K}^2}
{\rm~arccos}\left(-\frac{m_{\bar K}} {m_{\eta}} \right) \right.
\nonumber \\ && \qquad \qquad - \left.  \frac1{4\pi^2} m_{\bar K}^2 (m_\eta-m_{\bar
K}) \ln \frac{M_\Lambda}{m_\eta} \, \right\} \nonumber \\ &+& {\cal
O}\left( \frac1{f^2 M^3}, \frac1{M f^4} , \frac1{f^6} \right)
\label{eq:a-piN}
\end{eqnarray}
This expression can be mapped into the HBChPT result of
Ref.~\cite{Ka01}, since the trascendental function dependence is
exactly the same. This should be so because they build the
perturbative unitarity correction in HBChPT. Thus, one could identify
a linear combination of the leading order approximation of our
constants, with another linear combination of HBChPT constants. On the
other hand, if we assume the values of the best fit 
parameters, Eq.~(\ref{eq:lecs1}), for  $J_{\hat m\hat M}^0$
and $\Delta_{\hat m}^0$ and $\Delta_{\hat M}^0$ we obtain the
following numerical estimate  
\begin{eqnarray}
a_{\bar K N} = \underbrace{1.02}_{1/f^2} - \underbrace{0.53}_{1/f^2 M}+ 
\underbrace{0.28}_{1/f^2 M^2 } +  
\underbrace{\overbrace{1.00+\, {\rm i} \, 0.41}^{\pi \Sigma}-
\overbrace{1.25}^{\bar K N }+\overbrace{3.17}^{\eta \Lambda}}_{1/f^4}
+ \dots 
=3.69 + {~\rm i~} 0.41 \,~ {\rm fm} + \dots 
\label{eq:akbn_HB}
\end{eqnarray} 
where the contributions are separated according to the order in the
chiral expansion and also to the corresponding intermediate state.  As
we see, large cancellations at higher orders must take place to
obtain, after summing the whole series, our result in
Eq.~(\ref{eq:apin}). Note also that the real part is about twice and
with opposite sign as compared to the experimental result~\cite{Ma81}.
A similar situation occurs in HBChPT~\cite{Ka01}; the real part of the
scattering amplitude has opposite sign although similar magnitude to
the experimental number. Likewise, large cancellations have also been
noted, indicating a bad convergence rate in HBChPT. The fact that our
calculation, Eq.~(\ref{eq:akbn_HB}), gives a larger magnitude for
${\rm Re} \, a_{\bar K N} $ than in HBChPT reflects, in addition, a
bad convergence in the expansion (\ref{eq:hbexp}). This situation has
also been described in the coupled channel case $S=0$ sector,
\cite{JE01b} and seems a common feature of unitarization methods
\cite{pn00,ej00c,JE01}.

\section{Conclusions}
\label{sec:concl}

In this paper we have extended the Bethe-Salpeter formalism developed
in Ref.~\cite{JE01b} to study $s-$wave and $I=0$ meson-baryon
scattering up to 1.75 GeV in the strangeness $S=-1$ sector. We work on
a four dimensional two body channel space and the kernel of the BSE
takes into account CS constraints as deduced from the corresponding
effective Lagrangian.  The obtained $t$ matrix manifestly complies
with coupled channel unitarity and the undetermined low-energy
constants of the model have been fitted to data. The available direct
experimental information is limited to the $\pi \Sigma \to \pi \Sigma
$ mass spectrum, and the $K^- p \to \Lambda \eta $ total cross
section, for which errors are provided, and to $\bar K N \to \bar K N$
and $\bar K N \to \pi \Sigma $ scattering amplitudes of a partial wave
analysis, for which errors are guessed. Taking this into account, the
agreement with experiment is satisfactory. Besides, some predictions
for other cross sections, not yet measured, have been also given.  A
careful and detailed statistical study has been carried out, showing
that only seven parameters (LEC's) out of the starting twelve are
really independent. Thus, though our model has some more free
parameters than those required in Ref.~\cite{ORB02}, the description
of data achieved in our approach is superior to that of
Ref.~\cite{ORB02}. A similar situation was already found also in the
strangeness $S=0$ sector~\cite{JE01b}. According to previous
experience, the reduction of parameters is partly due to the
constraint of a well defined heavy baryon
limit~\cite{JE01b}. Likewise, crossing symmetry is expected to shed
more light on the number of independent LEC's. As we have argued,
matching to HBChPT calculations in the $S=-1$ and $I=0$ sector would
be the ideal way to map our LEC's into those stemming from an
effective chiral Lagrangian, but it is already known that
the chiral expansion fails~\cite{Ka01} to reproduce the $\bar K N$ scattering
length, due to the influence of the nearby subthreshold
$\Lambda(1405)$ resonance. 
Possibly  this could be overcome by 
properly accounting for the
singularity structure as suggested in~\cite{Meissner:1999vr}. 
We believe these points deserve a deeper
investigation.

We have undertaken a careful discussion on the analytical structure of
the scattering matrix amplitude in the complex $s-$plane, which
becomes mandatory in order to extract the features of the $S_{01}$
resonances. We have searched for poles in the second Riemann Sheet and
compared masses, widths and branching ratios to data. The
agreement is also quite satisfactory. In the resonance region our
unitary amplitude cannot be analyzed as a Breit-Wigner resonance due
to a sizeably non-diagonal background in coupled channel space. This,
in particular, prevents from a simple interpretation of branching
ratios. Although residues at the resonance poles are well and
unambiguously defined, the definition of branching ratios requires
special considerations and provisos, due to an ambiguous extrapolation
of the resonance contribution of the S-matrix from the pole to the
scattering line.  We have also illustrated that looking for a good
description of experimentally accessible data is not sufficient and
that, in some cases, it can be achieved at the expense of generating
non-physically acceptable poles in the first Riemann Sheet, which
influence on the scattering region is non-negligible. Thus, any fit to
data should be supplemented by this additional requirement of not
producing spurious singularities numerically relevant for the
description of scattering processes.

\section*{Acknowledgments}
We warmly thank E. Oset and A. Ramos for useful discussions. This
research was supported by DGES under contracts BFM2000-1326 and 
PB98-1367 and by the Junta de Andalucia.

\appendix

\section{Best fit results}\label{sec:app-stat}

The best fit ($\chi^2 / N_{\rm tot} = 0.93$) parameters are 

\begin{eqnarray} 
J_{\bar K N}&=-0.0186 &\pm 0.0010  \nonumber\\
J_{\pi \Sigma} &= 0.00796 &\pm 0.00061 \nonumber\\
J_{\eta \Lambda} &= 0.01264 &\pm 0.00021\nonumber\\
J_{K \Xi} &= -0.11936  &\pm 0.00018 \label{eq:lecs1}\\
\bar\Delta_{N}\equiv\Delta_{N} /(m_{\bar K}+M_N)^2 &= 0.01355 &\pm 0.00029  \nonumber\\
\bar\Delta_{\Sigma}\equiv\Delta_{\Sigma}/(m_\pi+M_\Sigma)^2 &= -0.00325  &\pm 0.00036 \nonumber\\
\bar\Delta_{\Lambda}\equiv\Delta_{\Lambda}/(m_\eta+M_\Lambda)^2  &= -0.00262 &\pm 0.00011 \nonumber
\end{eqnarray}
with fixed parameters
\begin{eqnarray}
\Delta_{\Xi} /(m_K+M_\Xi)^2 &=& -0.0035  \nonumber\\ 
\Delta_{\bar K} /(m_{\bar K}+M_N)^2&=& -0.034  \nonumber\\
\Delta_{\pi } /(m_\pi+M_\Sigma)^2&=& 0.060 \label{eq:lecs2}\\
\Delta_{\eta } /(m_\eta+M_\Lambda)^2 &=& 0.049  \nonumber\\ 
\Delta_{K}  /(m_K+M_\Xi)^2 &=& -0.26  \nonumber
\end{eqnarray} 
as explained in the main text. We assume that  the parameters of
Eq.~(\ref{eq:lecs1}) are Gaussian correlated, this is justified because they come
from a $\chi^2-$fit. 
To make any further statistical analysis of quantities derived from
the parameters above,
the corresponding 
covariance ($v$) and correlation ($c$) matrices are needed.
These matrices are defined as usual 
\begin{eqnarray}
v_{ij} &=& \left[\left(\frac{1}{2}\frac{\partial \chi^2}{\partial
    b_k\partial b_l}\right)^{-1}\right]_{ij} \nonumber\\
c_{ij} &=& v_{ij}/\sqrt{v_{ii} v_{jj}} , 
\end{eqnarray}
being $b_i$ any of the seven parameters $J's$ and $\Delta 's$ of
Eq.~(\ref{eq:lecs1}).  The errors, $\delta b_i$,  quoted in Eq.~(\ref{eq:lecs1})
are obtained from the diagonal elements of the covariance matrix ( $\delta b_i
=\sqrt{v_{ii}}$). 
Finally our estimate for the  correlation 
matrix reads:
\begin{eqnarray}
&{\scriptsize
\left(
\matrix{
&
J_{\bar K N}&
J_{\pi \Sigma} &
J_{\eta \Lambda} &
J_{K \Xi} &
\bar{\Delta}_{N} &
\bar{\Delta}_{\Sigma}&
\bar{\Delta}_{\Lambda}\cr
& & & & & & &\cr
\phantom{llll}J_{\bar K N}\phantom{llll} & 
\phantom{-}1.000 & & & & & &  \cr
J_{\pi \Sigma} &           -0.236 &\phantom{-}1.000 & &  & & & \cr
J_{\eta \Lambda} &
           -0.909 &\phantom{-}0.442 &\phantom{-}1.000  & & & & \cr
J_{K \Xi} &
 \phantom{-}0.569 &          -0.479 &          -0.530 &\phantom{-}1.000  & & & \cr
\bar{\Delta}_{N} &
           -0.830 &\phantom{-}0.228 &\phantom{-}0.702 &          -0.829 &\phantom{-}1.000 & & \cr
\bar{\Delta}_{\Sigma}&
 \phantom{-}0.294 &\phantom{-}0.608 &          -0.030 &\phantom{-}0.224 &-0.494 &\phantom{-}1.000  & \cr
\bar{\Delta}_{\Lambda} &
           -0.158 &          -0.501 &\phantom{-}0.087 &\phantom{-}0.613 &-0.336 &-0.051 &\phantom{-}1.000  \cr
}
\right)
}&
\label{eq:corr}
\end{eqnarray}
Despite the correlations in the above matrix are at most in modulus of
about $0.9$, the matrix has got an eigenvalue quite close to zero
(0.0025), which is a clear indication that one of the parameters might
still be redundant.

\section{Non-physically acceptable fits to data}\label{app:monster}

\begin{figure}[ht]
\vspace{-.75cm}
\begin{center}                                                                
\leavevmode
%\makebox[0cm]{\epsfbox{Rebdivrhor-kfr.ps}}
\makebox[0cm]{
\epsfysize = 150pt
\epsfbox{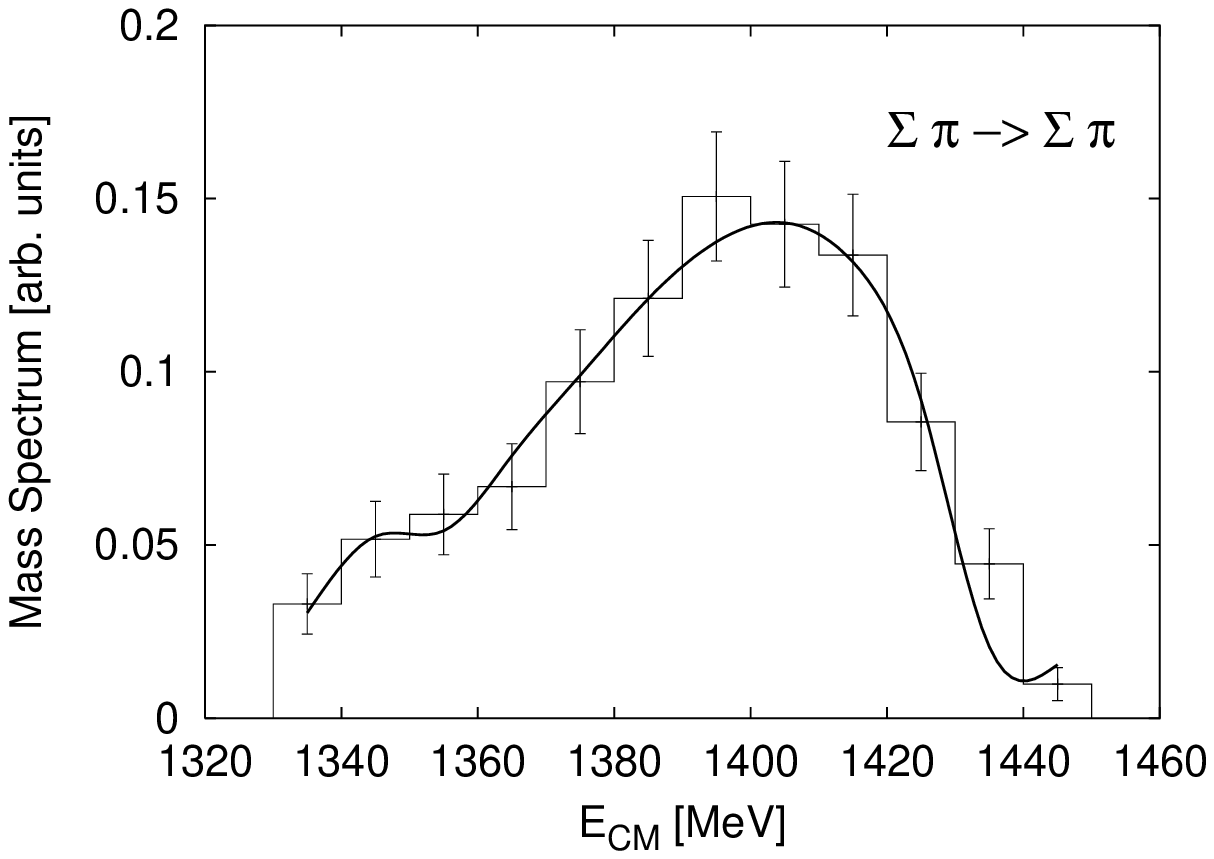}~
\epsfysize = 150pt
\epsfbox{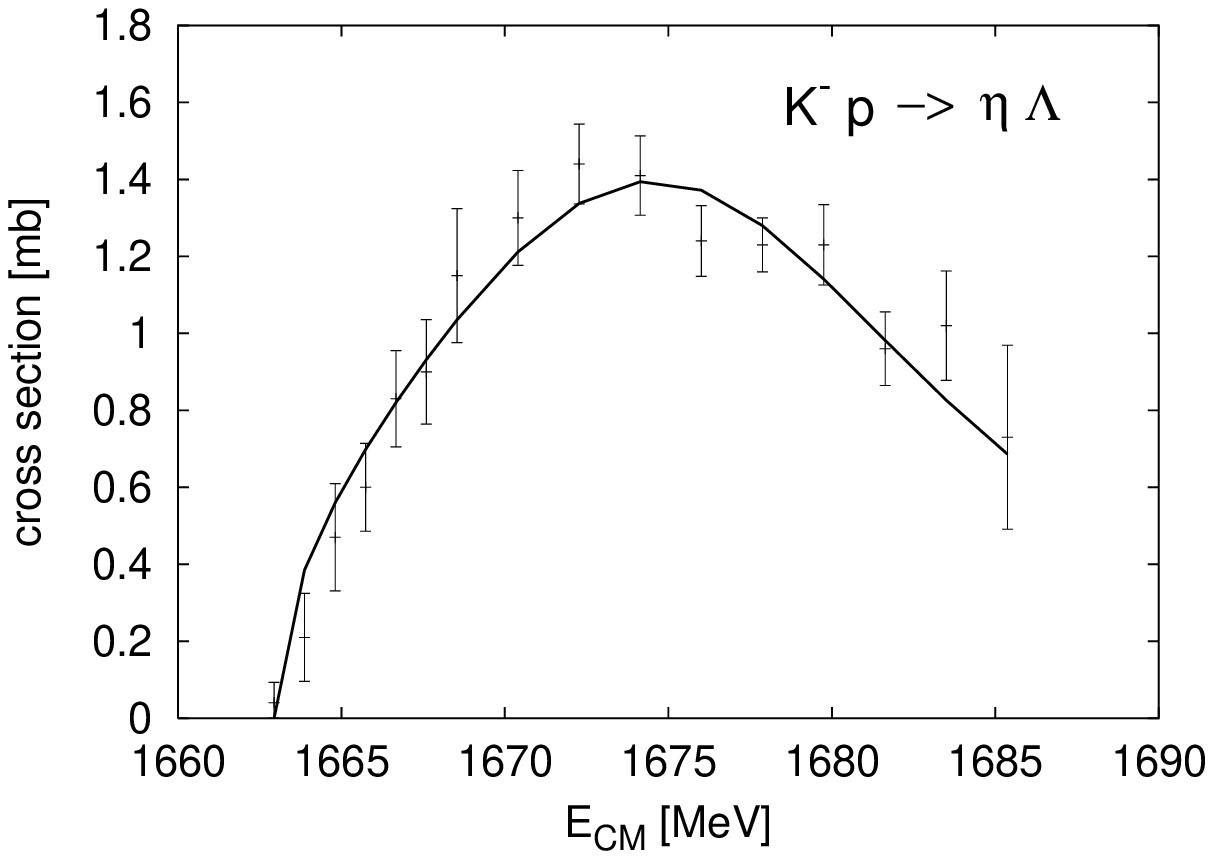}}
\end{center}
\vspace{.5cm}
\caption
{Same as in Fig.~(\protect\ref{fig:1}) for ``the non physically acceptable fit'' 
described in the Appendix~\protect\ref{app:monster}.}
\label{fig:1-m}
\end{figure}

\begin{figure}[ht]
\vspace{-.75cm}
\begin{center}                                                                
\leavevmode
\makebox[0cm]{
\epsfysize = 150pt
\epsfbox{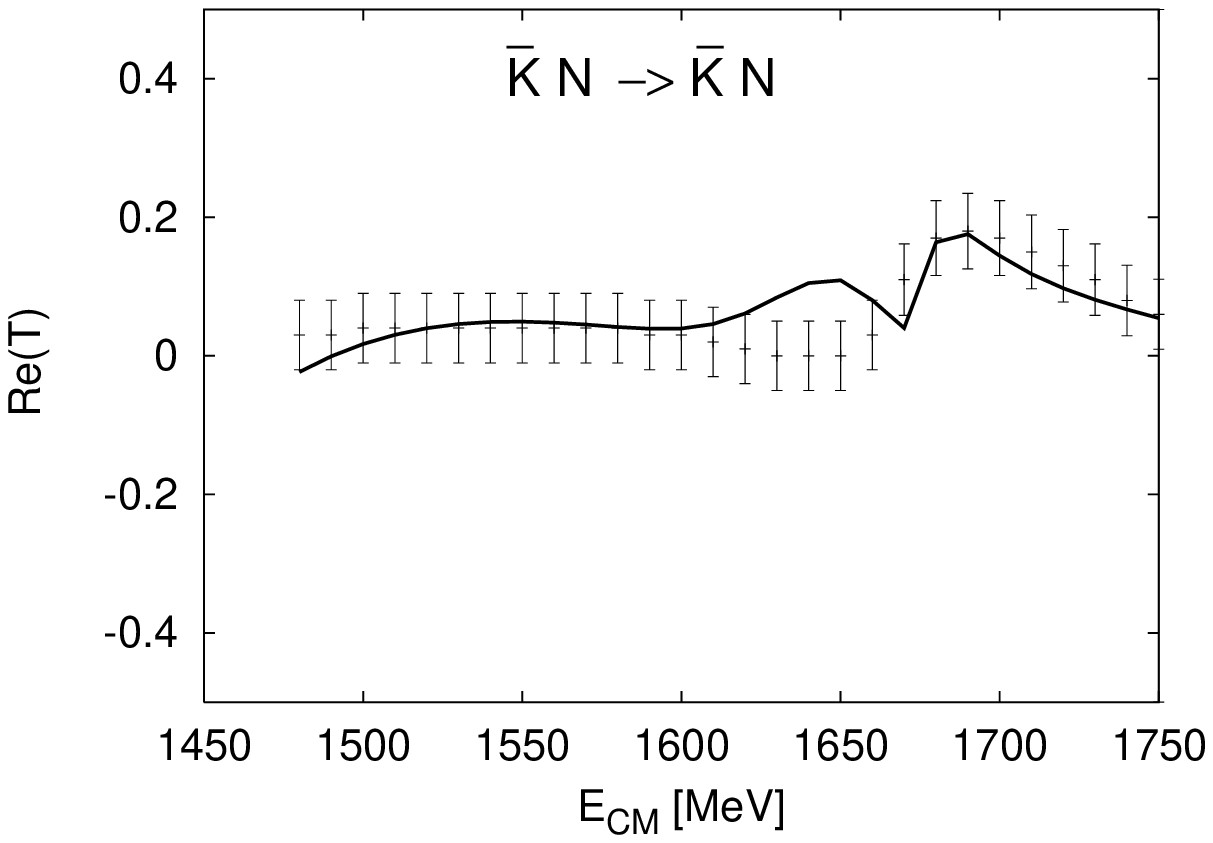}~
\epsfysize = 150pt
\epsfbox{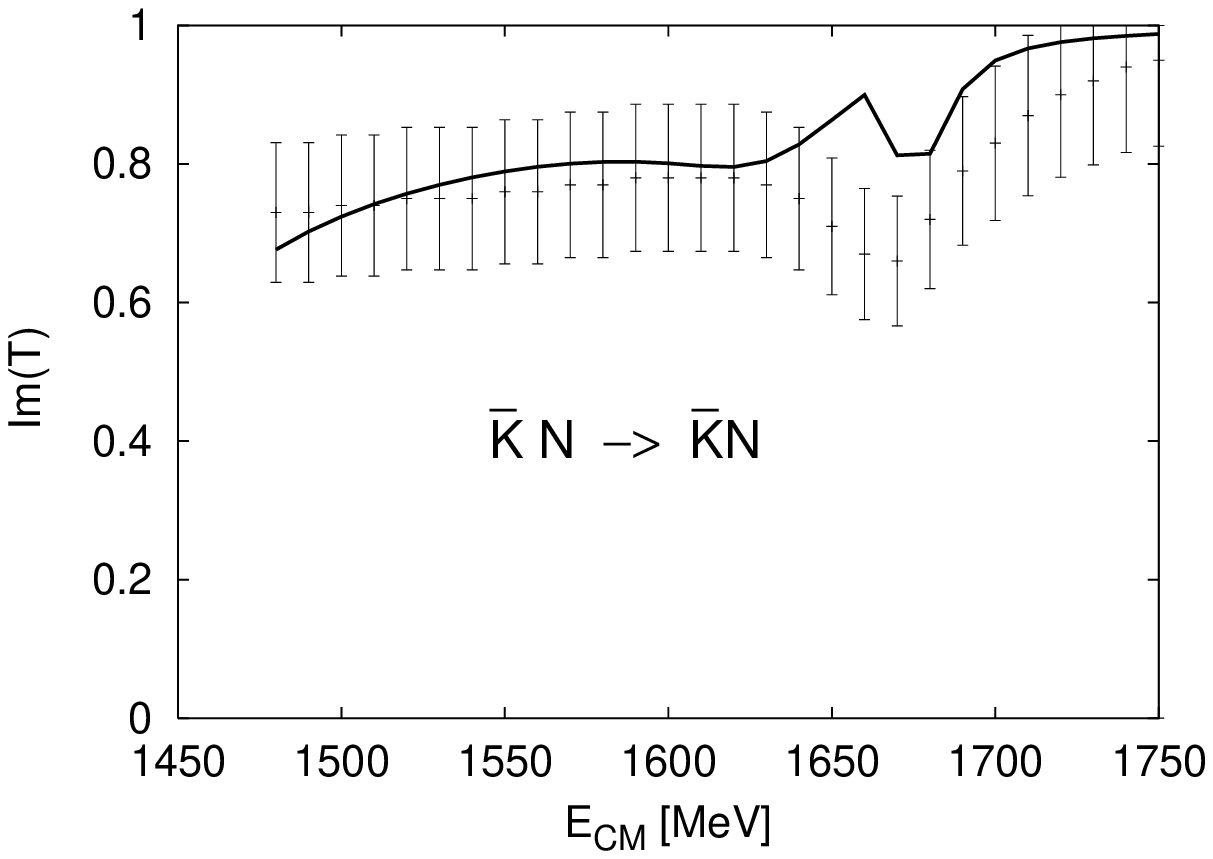}}
\end{center}
%\vspace{-.5cm}
\caption[pepe]{\footnotesize 
Same as in Fig.~(\protect\ref{fig:2}) for  ``the non physically acceptable fit'' 
described in the Appendix~\protect\ref{app:monster}.}
\label{fig:2-m}
\end{figure}

\begin{figure}[ht]
\vspace{-.75cm}
\begin{center}                                                                
\leavevmode
%\makebox[0cm]{\epsfbox{Rebdivrhor-kfr.ps}}
\makebox[0cm]{
\epsfysize = 150pt
\epsfbox{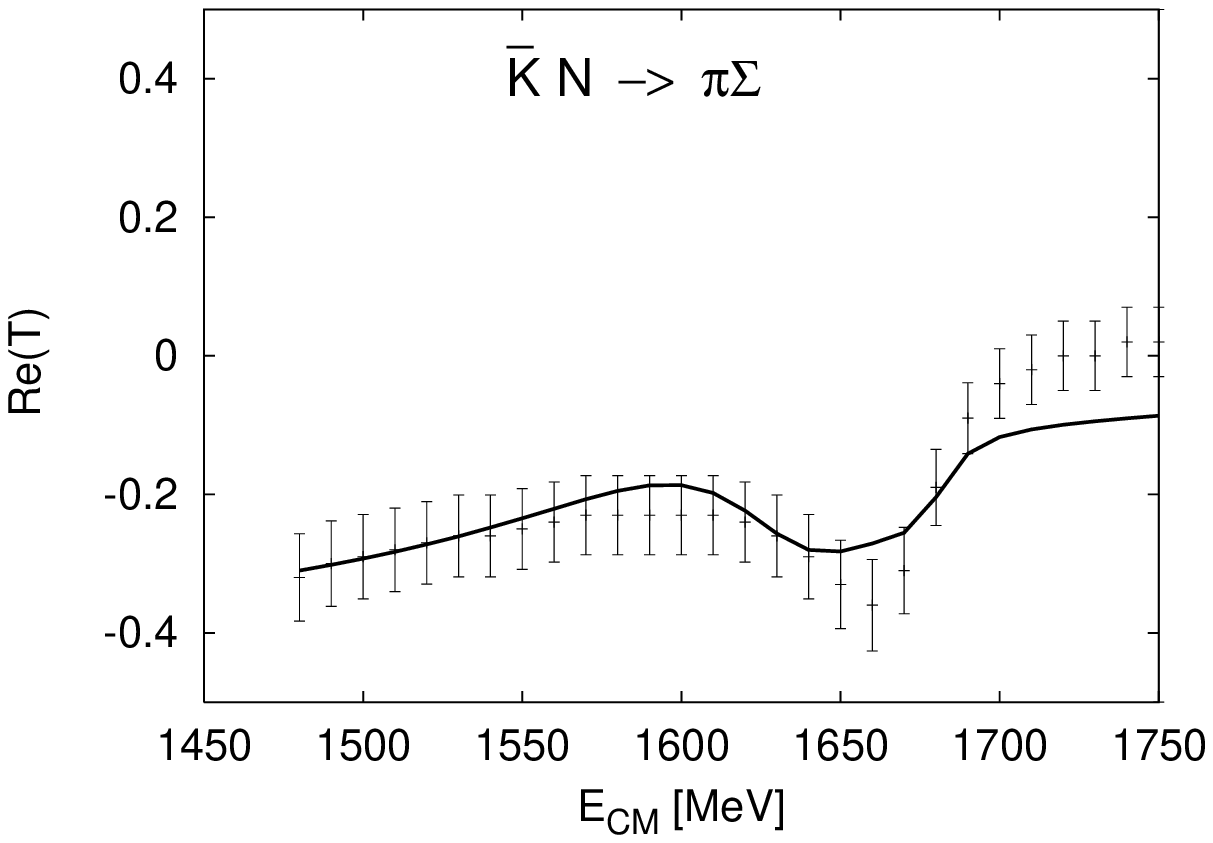}~
\epsfysize = 150pt
\epsfbox{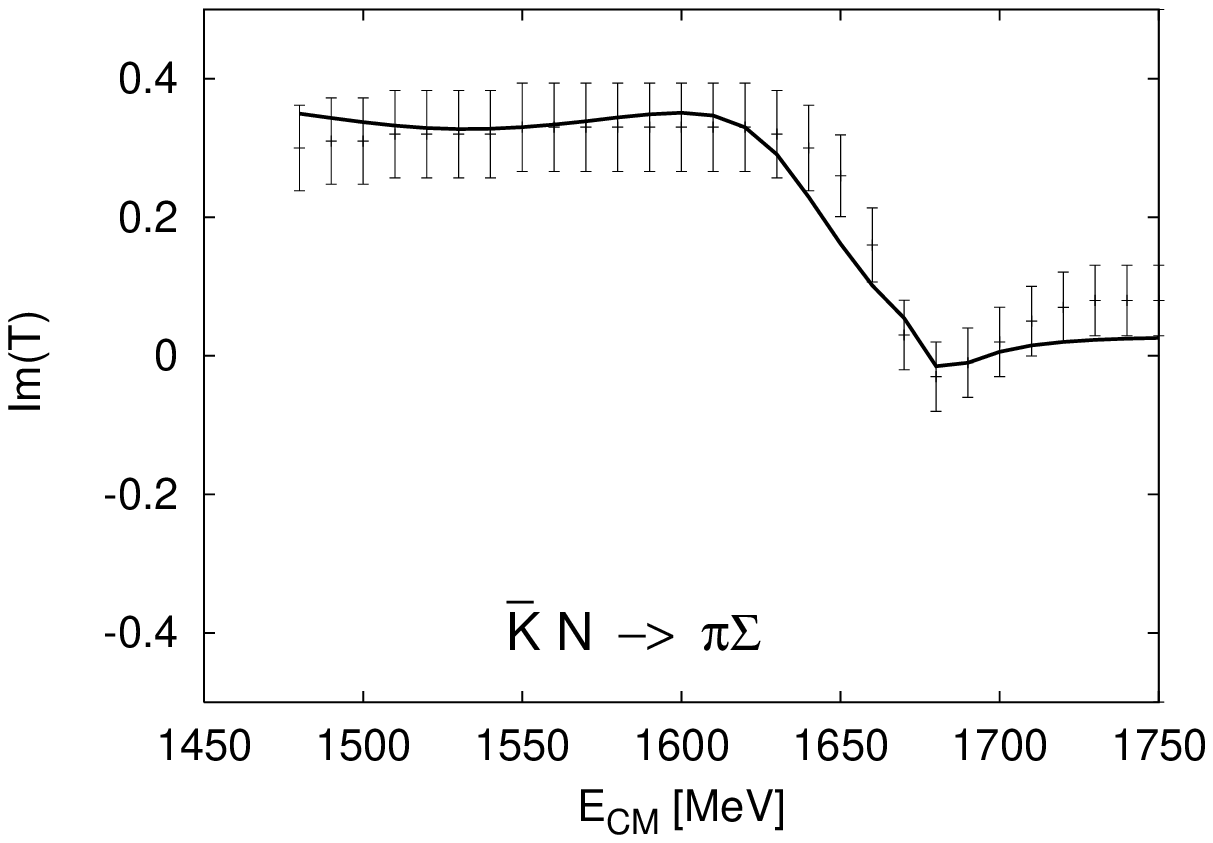}}
\end{center}
\vspace{.5cm}
\caption[pepe]{\footnotesize 
Same as in Fig.~\protect\ref{fig:2-m} for the 
inelastic channel $\bar K N \to \pi \Sigma$. }
\label{fig:3-m}
\end{figure}

\begin{figure}
\epsfysize = 350pt
\epsfbox{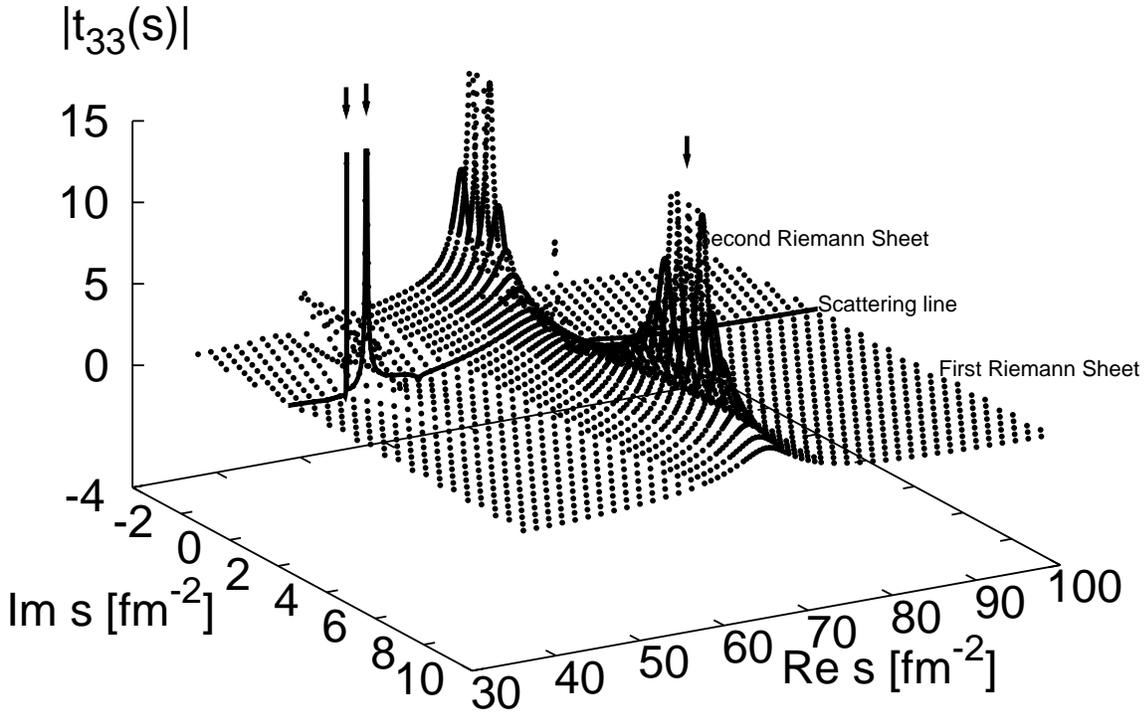}
%\epsfbox{h33-monster.ps}
%\centerline{\epsfig{figure=h33-7-3.ps,height=12cm,width=17cm}}
\caption[pepe]{\footnotesize 
Same as in Fig.~(\protect\ref{fig:riemman}) for a the non physically acceptable  fit to data
described in the Appendix\protect\ref{app:monster}.}
\label{fig:riemman-m}
\end{figure}

In Figs.~(\ref{fig:1-m}-\ref{fig:3-m}) we present the results of 
a fit to the data, which we will show is  not physically acceptable.
The overall description of data is remarkably good. This fit gives 
$\chi^2 / N_{\rm tot} = 0.69$ to compare with the value of 0.93 of the
best fit presented in the main text. 
%
%c-------------------------------------------
\vspace*{.5cm}
%
%c--------------------------------------------------------------
%c-------------------------------------------
%
\vspace*{.5cm}
%
%c--------------------------------------------------------------
%c-------------------------------------------
%
\vspace*{.5cm}
%
%c--------------------------------------------------------------
%
However, when looking at the $t(s)$ matrix in the $s-$complex plane
(Fig.~(\ref{fig:riemman-m})), one realizes that there exists a
proliferation of poles, some of them unphysical and others with no
experimental counterparts. In the first Riemann Sheet we find at least
two poles. The first one is located at $s = M^2 - {\rm i} M \Gamma$
with $M = 1606$~MeV and $\Gamma = 153$~MeV. This pole is close to the
$s$ real axis and produces visible effects on the scattering line not 
only for the $t_{\eta\Lambda\to\eta\Lambda}$ entry shown in the
figure, but also for all $i\to j$ channels.  Indeed, as we showed in
Subsect.~\ref{sec:reso}, our preferred fit has also a similar
unphysical pole but significantly farther ($\Gamma=631$~MeV) from the
real axis, and therefore with a tiny influence on the physical
scattering. Since causality imposes the absence of poles in the
physical Sheet~\cite{Ma58}, the existence of such a pole affecting the
scattering line invalidates the description (
Figs.~(\ref{fig:1-m}-\ref{fig:3-m})) presented in this Appendix,
despite of its quality. Besides in the first Sheet, there exists a
pole on the real axis ($\sqrt{s}= 1307 $ MeV) and below the first
threshold, which would correspond to a bound state, stable under
strong interactions. Such a state has the same quantum numbers as the
$\Lambda (1115)$ and it should show up in all reactions where the
latter one is produced.

On the other hand, in the second Sheet there exist now four
poles. Three of them are similar to those presented in
Subsect.~\ref{sec:reso}, though the one placed around 1370 MeV
(Eq.~(\ref{eq:1300})) is now almost a factor two narrower (it is
located at $M_R = 1392$~MeV and $\Gamma_R = 120$~MeV).  In addition
there exists a new resonance $M_R = 1343$~MeV and $\Gamma_R =
0.18$~MeV which is responsible of the high peak at the beginning of
the scattering line in Fig.~\ref{fig:riemman-m} and of the existing
bump between 1330 and 1360 MeV in the $\pi \Sigma$ mass spectrum of
Fig.~\ref{fig:1-m}. As far as we know, there are no other independent
indications of the existence of this extremely narrow resonance.

  We have presented the results of this
`` non physically acceptable  fit'' to stress that, in order to be
sure of having a good approach to the $t$ matrix of a
given physical system, one should {\it not only} look at the $t$
matrix and related observables (cross sections, ...) at the physical
scattering line (real $s$ values), but one should {\it also} study its
behavior on the $s-$complex plane, both in the second Riemann Sheet to
find the resonances and in the first Riemann Sheet to be sure of
avoiding pathological behaviors as the one illustrated in
Fig.~{\ref{fig:riemman-m}}.

\end{document}